\documentclass{elsart}

\newcounter{bla}
\newenvironment{refnummer}{%
\list{[\arabic{bla}]}%
{\usecounter{bla}%
 \setlength{\itemindent}{0pt}%
 \setlength{\topsep}{0pt}%
 \setlength{\itemsep}{0pt}%
 \setlength{\labelsep}{2pt}%
 \setlength{\listparindent}{0pt}%
 \settowidth{\labelwidth}{[9]}%
 \setlength{\leftmargin}{\labelwidth}%
 \addtolength{\leftmargin}{\labelsep}%
 \setlength{\rightmargin}{0pt}}}
 {\endlist}

\usepackage{epsf}
\usepackage{epsfig}
\usepackage{graphicx}
\usepackage{subfigure}
\usepackage{xspace}
\usepackage{multirow}

\usepackage [usenames,dvipsnames]{color}
\newcommand{\sla}{\kern -5.4pt /}
\newcommand{\Dir}{\kern -6.4pt\Big{/}}
\newcommand{\Dirin}{\kern -10.4pt\Big{/}\kern 4.4pt}
\newcommand{\DDir}{\kern -7.6pt\Big{/}}
\newcommand{\DGir}{\kern -6.0pt\Big{/}}

\newcommand{\be}{\begin{equation}}
\newcommand{\ee}{\end{equation}}
\newcommand{\bea}{\begin{eqnarray}}
\newcommand{\eea}{\end{eqnarray}}
\newcommand{\beanon}{\begin{eqnarray*}}
\newcommand{\eeanon}{\end{eqnarray*}}
\newcommand{\ba}{\begin{array}}
\newcommand{\ea}{\end{array}}
\newcommand{\bd}{\begin{description}}
\newcommand{\ed}{\end{description}}
\newcommand{\bi}{\begin{itemize}}
\newcommand{\ei}{\end{itemize}}
\newcommand{\ben}{\begin{enumerate}}
\newcommand{\een}{\end{enumerate}}
\newcommand{\bc}{\begin{center}}
\newcommand{\ec}{\end{center}}

\newcommand{\vsk}{\vskip 10 pt\noindent}

\newcommand{\epem}{\mbox{${\mathrm e}^+{\mathrm e}^-$}\xspace}

\newcommand{\toptop}{\mbox{${\mathrm t} \bar {\mathrm t}$}\xspace}

\newcommand{\VVL}{\mbox{${\mathrm V_L}{\mathrm V_L}$}\xspace}

\newcommand{\VV}{\mbox{${\mathrm V}{\mathrm V}$}\xspace}
\newcommand{\VVjj}{\mbox{${\mathrm V}{\mathrm V}+2\,jets$}\xspace}

\newcommand{\PW}{\mathrm W}
\newcommand{\PZ}{\mathrm Z}

\newcommand{\ordEW}{$\mathcal{O}(\alpha_{\scriptscriptstyle EM}^6)$\xspace}
\newcommand{\ordQCD}{$\mathcal{O}(\alpha_{\scriptscriptstyle EM}^4
  \alpha_{\scriptscriptstyle S}^2)$\xspace}
\newcommand{\ordQCDsq}{$\mathcal{O}(\alpha_{\scriptscriptstyle EM}^2
  \alpha_{\scriptscriptstyle S}^4)$\xspace}

\newcommand{\eqn}[1]{Eq.(\ref{#1})}

\newcommand{\fig}[1]{Fig.~\ref{#1}}

\newcommand{\Phase}{{\tt PHASE}\xspace}
\newcommand{\Phact}{{\tt PHACT}\xspace}
\newcommand{\Phantom}{{\tt PHANTOM}\xspace}
\newcommand{\Pythia}{{\tt PYTHIA}\xspace}

\newcommand{\LHA}{Les Houches Accord\xspace}
\newcommand{\veg}{{\tt VEGAS}\xspace}

\def\pl #1 #2 #3 {{\it Phys.~Lett.} {\bf#1} (#2) #3}   
\def\np #1 #2 #3 {{\it Nucl.~Phys.} {\bf#1} (#2) #3}
\def\zp #1 #2 #3 {{\it Z.~Phys.} {\bf#1} (#2) #3}
\def\pr #1 #2 #3 {{\it Phys.~Rev.} {\bf#1} (#2) #3}
\def\prep #1 #2 #3 {{\it Phys.~Rep.} {\bf#1} (#2) #3}
\def\prl #1 #2 #3 {{\it Phys.~Rev.~Lett.} {\bf#1} (#2) #3}
\def\intj #1 #2 #3 {{\it Int. J. Mod. Phys.} {\bf#1} (#2) #3}
\def\mpl #1 #2 #3 {{\it Mod.~Phys.~Lett.} {\bf#1} (#2) #3}
\def\rmp #1 #2 #3 {{\it Rev. Mod. Phys.} {\bf#1} (#2) #3}
\def\cpc #1 #2 #3 {{\it Comp. Phys. Commun.} {\bf#1} (#2) #3}
\def\epj #1 #2 #3 {{\it Eur. Phys. J.} {\bf#1} (#2) #3}
\def\jhep #1 #2 #3 {{\it JHEP} {\bf#1} (#2) #3}

\begin{document}
\begin{frontmatter}

\title{  \Phantom : a Monte Carlo event generator for six parton                final states at high energy colliders.}

\author[a]{Alessandro Ballestrero},
\author[c]{Aissa Belhouari},
\author[a,b]{Giuseppe Bevilacqua}
\author[a,b]{Vladimir Kashkan}
\author[a,b]{Ezio Maina\thanksref{author}}

\thanks[author]{Corresponding author.  Email: {\tt maina@to.infn.it}  }

\address[a]{INFN, Sezione di Torino, Italy}
\address[b]{Dipartimento di Fisica Teorica, Universit\`a di Torino, Italy}
\address[c]{The Abdus Salam International Center for Theoretical Physics,
 Trieste, Italy}

\begin{abstract}
 \Phantom is a tree level Monte Carlo  for  six
parton final states at proton--proton, proton--antiproton and electron--positron
collider at \ordEW and \ordQCD including
possible interferences between the two sets of diagrams.
This   comprehends  all purely electroweak contributions as 
well as all contributions  with one virtual or two external gluons.
It can generate unweighted events for any set of processes and it is interfaced
to parton shower and hadronization packages via the latest Les Houches Accord
protocol. It can be used to analyze the physics of   boson boson scattering,
Higgs boson production in boson boson fusion, $t\bar t$ and three boson  production.

\begin{flushleft}
\end{flushleft}


\end{abstract}

\end{frontmatter}


{\bf PROGRAM SUMMARY/NEW VERSION PROGRAM SUMMARY}

\begin{small}
\noindent
{\em Manuscript Title:}  \Phantom : a Monte Carlo event generator for six 
         parton final states at high energy colliders             \\
{\em Authors:} A.~Ballestrero, A.~Belhouari, G.~Bevilacqua, V~Kashkan,
E.~Maina                                                          \\
{\em Program Title:}  \Phantom (V. 1.0)                             \\
{\em Journal Reference:}                                      \\
{\em Catalogue identifier:}                                   \\
{\em Licensing provisions:}   none                                \\
{\em Programming language:} Fortran 77                             \\
{\em Operating system:} UNIX, LINUX                               \\
{\em RAM required to execute program with typical data:} 500MB         
{\em Classification:}  11.1 General, High Energy Physics and Computing     \\
{\em External routines/libraries:}   LHAPDF (Les Houches Accord PDF Interface),
CIRCE (beamstrahlung for \epem ILC collider). \\
{\em Nature of problem:}\\Six fermion final state processes have become 
important with the increase of collider  energies  and are essential
for the study of top, higgs and electroweak symmetry breaking physics
 at high energy colliders. 
Since thousands of Feynman diagrams contribute in a single process and 
events corresponding to hundreds of different final states need to be 
generated, a fast and stable calculation is needed.
   \\
{\em Solution method:}\\ 
  \Phantom is a tree level Monte Carlo  for  six
parton final states at proton--proton, proton--antiproton and 
electron--positron collider. It computes all amplitudes at \ordEW and 
\ordQCD including possible interferences between the two sets of diagrams.
The matrix elements are computed with the helicity formalism implemented
in the program \Phact [1]. 
The integration makes use of an iterative-adaptive multichannel 
method which, relying on adaptivity, allows to use only few channels per
process. Unweighted event generation can be performed for any set of processes 
and it is interfaced to parton shower and hadronization packages via the 
latest Les Houches Accord protocol.
   \\
{\em Restrictions:}\\ All Feynman diagrams are computed al LO.
   \\
{\em Unusual features:}\\ \Phantom is written in Fortran 77 but it makes
uses of structures. The g77 compiler cannot compile it as it does not recognize
the structures. The vIntel, Portland Group,  True64 HP  fortran77 or fortran90
compilers have been tested and can be used.
   \\
{\em Typical running time:}\\ A few hours for a cross section integration
of one process at per mille accuracy. One hour for  one thousand unweighted 
events.   
   \\
{\em References:}
\begin{refnummer}
\item Reference 1 A.~Ballestrero and E.~Maina, \pl B350 1995 225 ,
[hep-ph/9403244]. 
A.~Ballestrero, {\tt PHACT 1.0 - \it Program for Helicity Amplitudes Calculations 
with Tau matrices} [hep-ph/9911318] in {\it 
Proceedings of the 14th International Workshop on High Energy Physics 
and Quantum Field Theory (QFTHEP 99)}, 
B.B.~Levchenko and V.I.~Savrin  eds. (SINP MSU Moscow), pg. 303.         
\end{refnummer}
\end{small}

\newpage


\hspace{1pc}
{\bf LONG WRITE-UP}

\section{Introduction}
\label{sec:intro}
Monte Carlo event generators are essential tools for the comparison of theory
and experiment in High Energy Physics. 
In this paper we present \Phantom a new event generator which is capable of
simulating any set of reactions with six partons in the final state at $pp$,
$p\overline{p}$ and $e^+e^-$ colliders at \ordEW and \ordQCD including
possible interferences between the two sets of diagrams.
This includes all purely electroweak contributions as well as all contributions
with one virtual or two external gluons. 
The relevance of six fermion final states both at the LHC and the ILC is well known and
will be further discussed in Sect. \ref{sec:physproc}. The signal of processes
like  Higgs boson production in vector boson fusion with a Higgs
decaying in two bosons,  boson boson scattering and 3-vector boson production
are all described by 6 fermion final states at  \ordEW .
Top-antitop pair production is also described by a six fermion final state and
its signal is at \ordEW for the ILC but at hadron colliders the main contribution comes
from \ordQCD .

As far as the irreducible backgrounds to the above mentioned signal
is concerned,  \ordQCD reactions encompass the full lowest order QCD background
for
six-parton studies of final states with four leptons at hadron colliders. 
An example is \VV fusion with fully leptonic decays of the
intermediate vector bosons. 
On the contrary, semileptonic channels get additional contributions from 
\ordQCDsq  processes which are responsible for the $V+4 \,jets$ background. 
The latter must be included as well in a complete analysis and can be covered by
other Monte Carlo event generators, such as \texttt{AlpGen} \cite{Mangano:2002ea}
or \texttt{MadEvent} \cite{Madgraph}. 
It should be noticed nevertheless that these contributions have quite 
different kinematical features with respect to the scattering signature, 
therefore they are expected to be suppressed by means of appropriate 
selection cuts.

At $e^+e^-$ colliders there is no QCD background for six and four 
lepton final states
and \ordQCD reactions describe
completely the QCD background to final states with two
leptons. Only for fully hadronic final states the set of reactions presently
available in
\Phantom needs to be complemented by other generators.

\Phantom employs exact tree matrix elements.  
There are issues which cannot be tackled without a complete
calculation, like exact spin correlations between the decays of different heavy
particles, the effect of the non resonant background, the relevance of the
offshellness of boson decays, the question of interferences between
different subamplitudes.
Without a full six parton computation it will be impossible to
determine the accuracy of approximate estimates. In Ref.\cite{Accomando:2005hz}
the complete calculation for $PP\rightarrow 4j\mu\bar\nu_\mu$\xspace at
\ordEW 
has been compared at length with a production times decay approach,
showing differences of the order of 10--20\% in some important regions of
phase space. The reliability of the Equivalent Vector Boson
Approximation (EVBA) \cite{EVBA} in the context of vector boson
scattering has been critically examined in \cite{Accomando:2006mc}.

\Phantom is an example of a {\it dedicated} event generator which describes
a predefined set of reactions striving for maximum speed and efficiency.
Other recent examples of dedicated programs for LHC physics are 
{\tt Toprex} \cite{Slabospitsky:2002ag},
which provides the matrix elements for several reactions related to top
production, {\tt Alpgen} \cite{Mangano:2002ea} and {\tt Gr@ppa} \cite{grappa}.
{\tt Alpgen} can simulate a large number of processes with electroweak bosons
and heavy quarks plus jets. When more than one electroweak boson are present in
the final state they are considered to be on shell.
{\tt Gr@ppa} provides event generators for V (W or Z)+ jets ($\leq$ 4 jets),
VV + jets ($\leq$ 2 jets) and QCD multi--jet ($\leq$ 4 jets) processes.
 
Dedicated generators aimed at future $e^+e^-$ colliders are
{\tt Lusifer} \cite{Dittmaier:2002ap}, which covers all reactions with six massless
fermions in the final state at \ordEW and \ordQCD, 
{\tt SIXFAP} \cite{sixfap}, which covers all reactions with six
fermions in the final state at \ordEW including mass effects, 
{\tt SIXPHACT} \cite{sixphact}, which was among the first MC to compute
six fermion final states but was limited to processes with one final neutrino,
{\tt eett6f} \cite{eett6f}, which simulates only processes related to
top--antitop production
and {\tt SIXRAD}\cite{sixrad}, which deals with all six jet final states at 
\ordQCDsq .

The Monte Carlo's just mentioned show that six fermion physics has been
already investigated 
at the ILC since a long time  and  they have been used for several 
phenomenological studies. \Phantom is the first complete dedicated Monte Carlo
which extends this kind of physics to hadron colliders. 

The complementary approach is 
given by {\it multi-purpose} programs for the automatic generation of any 
user-specified parton level process. The following codes for multi-parton 
production are available: {\tt Amegic-Sherpa} \cite{Amegic}, {\tt CompHEP} 
\cite{Comphep}, {\tt Grace} \cite{Grace}, {\tt Madevent} 
\cite{Madgraph}, {\tt Phegas \& Helac} \cite{Phegas}, 
{\tt O'Mega \& Whizard} \cite{Omega}.
{\tt CompHEP} is limited to processes with at most eight external particles
but since it computes matrix elements squared  instead of helicity matrix
elements it is extremely slow when a large number of particles is involved.
{\tt Grace} is a framework for generating single process matrix elements which
can be interfaced to integration and event generator modules.
The remaining packages, {\tt Amegic-Sherpa}, {\tt Madevent},
{\tt Phegas \& Helac} and {\tt O'Mega \& Whizard}
are complete event generators which automatically generate
the amplitudes, produce the mappings for integration over phase space,
compute cross sections and generate unweighted events. While processes with six
particles in the final states can be tackled with these generators, see for
instance \cite{Gleisberg:2003bi}, they are at
the limit of their practical capabilities when the unweighted generation of
several hundreds of this kind of processes is involved, as it is needed for
LHC physics.

\Phantom profits from the experience obtained with \Phase \cite{ref:Phase}
which could only simulate all electroweak processes of the type
$PP\rightarrow 4q + l \nu_{l}$ at the LHC. The two codes
share a number of key features. The matrix elements are computed with 
the use of the modular helicity formalism of Refs.\cite{method,phact}
which is well
suited to compute in a fast and compact way parts of diagrams of increasing 
size, and recombine them later to obtain the final set.
The integration strategy merges the 
multichannel approach \cite{multichannel} with the adaptivity of 
\veg \cite{vegas}. This results in generators which adapt
to different kinematical cuts and peaks with good 
efficiency, using only few channels per process.
Both codes employ the {\it one-shot} method developed for
{\tt WPHACT} \cite{wphact}, 
and used for four-fermion data analyses at LEP2. In this running 
mode, all processes are simultaneously generated in the 
correct relative proportion for any set of experimental cuts. 

The paper is organized as follows: in Sect.\ref{sec:physproc} we examine the physical content
of the processes which can be described by \Phantom. Then Sect.\ref{sec:descript} describes
the features of the program, the way amplitudes are calculated, phase spaces are implemented
and integration is performed. The following section is dedicated to the modes of operating
the program and finally Sect.\ref{sec:example} reports an example of the phenomenological
studies which can be performed using \Phantom .

\section{Physical processes}
\label{sec:physproc}
Processes  with six partons in the final state will be central to the
physics program at next generation accelerators, the LHC
\cite{ATLAS-TDR,CMS-TDR} and the ILC \cite{Djouadi:2007ik}.
They include Higgs boson production in vector boson fusion
followed by Higgs decay to $WW$ and $ZZ$, boson boson scattering processes,
top-antitop pair production, three vector boson production.

The Standard Model (SM) provides the simplest and most economical
explanation of Electro--Weak Symmetry Breaking (EWSB) in terms of a single Higgs
doublet. The search for the Higgs sector and its investigation
will have the highest priority for all LHC experiments
\cite{Atlas_HinWW,CMS:Higgs1,Atlas_HinVBF}.
Detailed reviews and extensive bibliographies can be found in 
Refs.\cite{HiggsLHC,djouadi-rev1,Houches2003}.
In the SM the Higgs is essential to the
renormalizability of the theory and is also crucial to 
ensure that perturbative unitarity bounds are not violated in high energy
reactions. In particular, if no Higgs is present in the spectrum,
longitudinally polarized vector bosons
interact strongly at high energy,
violating perturbative unitarity at about one TeV \cite{reviews}.
If, on the contrary, a relatively light Higgs exists then
they are weakly coupled at all energies.
These processes have been scrutinized since
a long time, going from the pioneering works in \cite{history1,history2}, which
address boson boson scattering on a general ground, to the more recent papers in
\cite{history3,history4} focused on the extraction of signals of vector boson
scattering at the LHC. In the last few years QCD corrections to boson boson 
production via vector boson fusion \cite{JagerOleariZeppenfeld} and to
three vector boson production \cite{Pittau_3V} at the LHC have been computed.
The size of the corrections depends quite strongly on the particular process
under consideration: while in the boson fusion case the corrections are below
10\%, for the three boson production case they are in the 70 to 100\% range.

Top pair production with its large cross section and large signal to background
ratio will play a central role since the early days of data taking at the LHC.
The measurement accuracy will quickly be dominated by systematic effects, in
particular by $b$--jet energy scale uncertainties, while statistical errors will
be negligible. The wealth of available channels will provide feedback on the
detector performance and assist in detector calibration
\cite{ATLAS-TDR,CMS-TDR}.
A total error on the top mass of about 2 GeV is expected to be feasible.
Exploiting the high statistics it will be possible to search for new physics
signatures in top production and decay.
Spin correlations will be measured to a precision of a few percent.
Finally, a good understanding of top physics will be essential since top
production will be a major source of background for all investigations of
processes involving
high $p_T$ $W$'s, like boson boson scattering and many new physics searches. 

\Phantom is intended in particular for studies of Higgs, boson boson 
scattering and top physics in six fermion final states. It can compute all
processes of this kind and does not make use of any production times decay
(narrow width) or equivalent vector boson approximation. 
This implies that every single diagram contributing to
a definite final state is included and not only the resonant ones.
So for instance if one studies a vector boson fusion process not only the diagrams
reported in Fig.1 will be considered but also, when appropriate, all other
diagrams of the types described in Figs. 2,3,4. In the same way when studying
top antitop production not only the first two diagrams on the left of Fig.~4
will be considered  but all the others schematically described in the
rest of Fig. 4 and also in the other 3 figures. Moreover all diagrams which
do not have two resonant bosons in the final state contribute as well.
We have already discussed in some particular cases \cite{sixphact,Accomando:2005hz,
Accomando:2006mc}  the differences between a 
complete calculation and an approximated one. These depend on the particular physical 
study at hand and on the applied cuts. But a complete calculation is in any case 
an important tool to quantify the possible discrepancies.

\begin{figure}
\begin{center}
\mbox{\includegraphics*[width=12.cm]{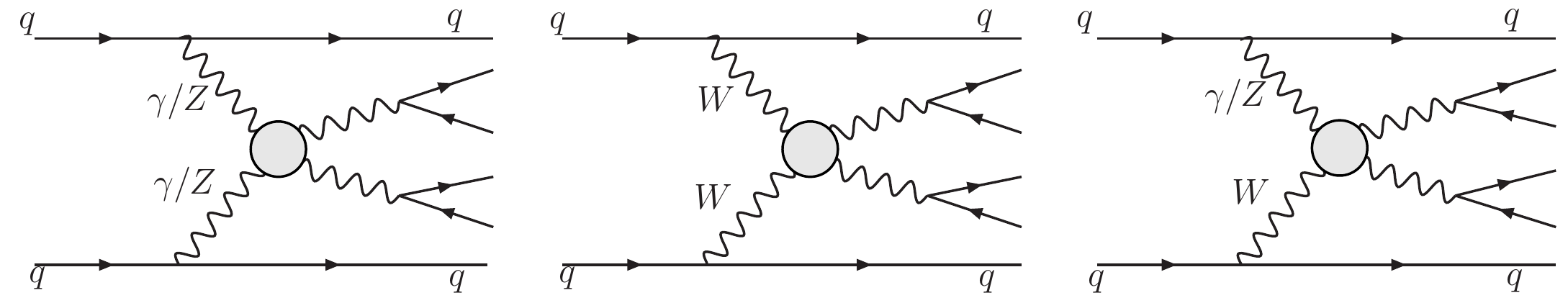}}
\caption{ Vector boson fusion processes.}
\label{VV-diag}
\end{center}
\end{figure}

\begin{figure}
\begin{center}
\mbox{\includegraphics*[width=14.cm]{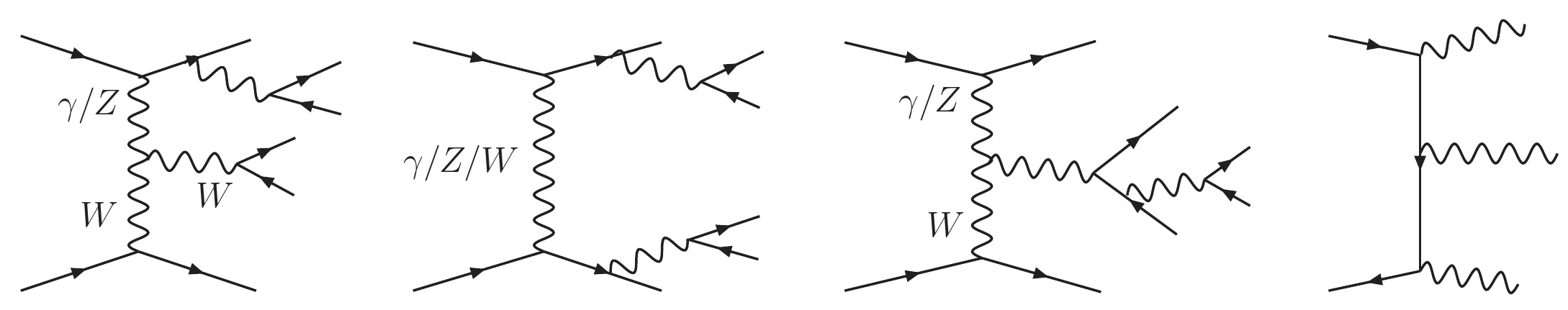}}
\caption{ Non fusion and non doubly resonant two vector boson production.      Three vector boson production.}
\label{nonreso-diag}
\end{center}
\end{figure}

\begin{figure}
\begin{center}
\mbox{\includegraphics*[width=9.cm]{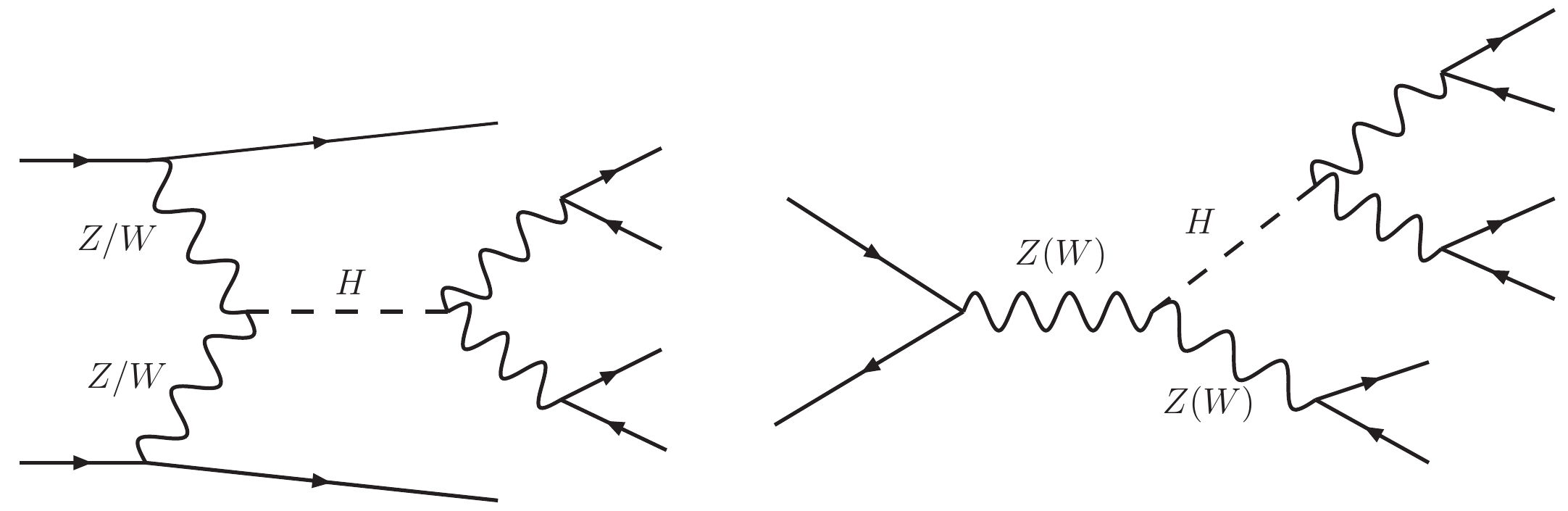}}
\caption{Higgs boson production via vector boson fusion
  and Higgsstrahlung.}
\label{higgs-diag}
\end{center}
\end{figure}

\begin{figure}[h!tb]
\centering
\includegraphics*[width=0.26\textwidth,height=2.8cm]{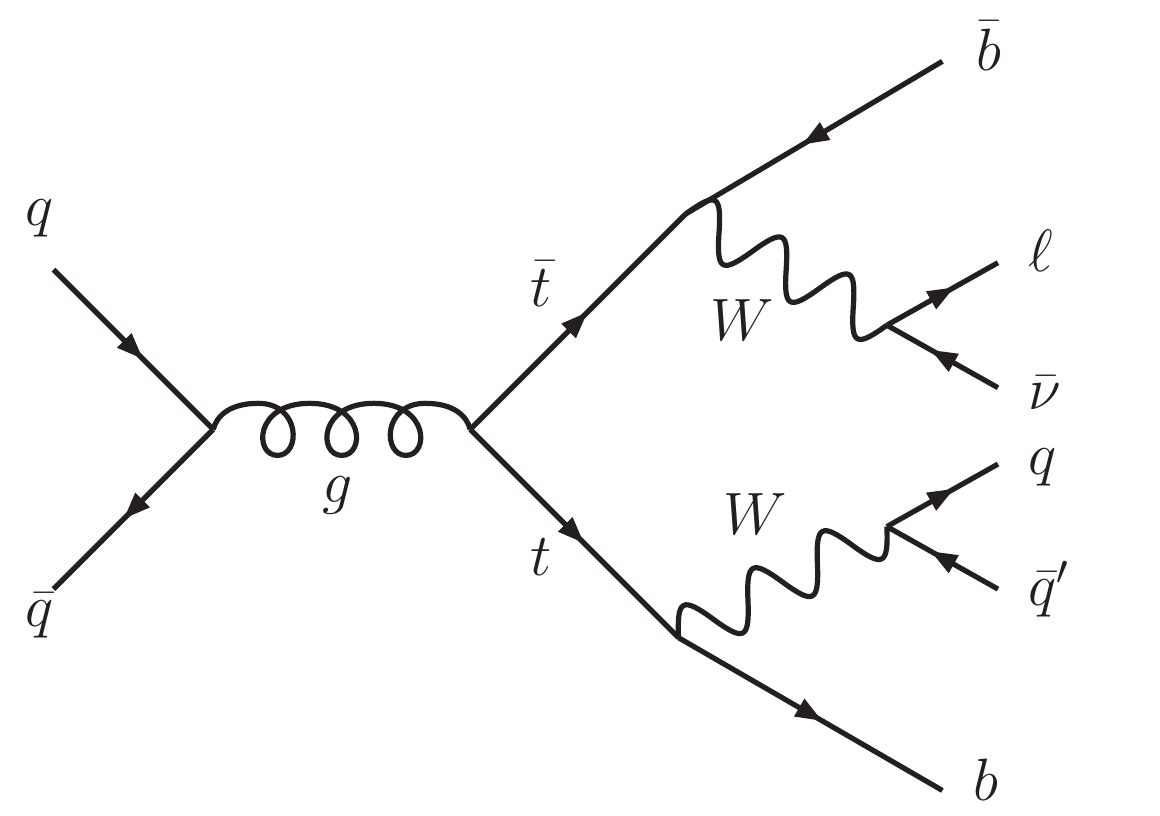}
\includegraphics*[width=0.26\textwidth,height=2.8cm]{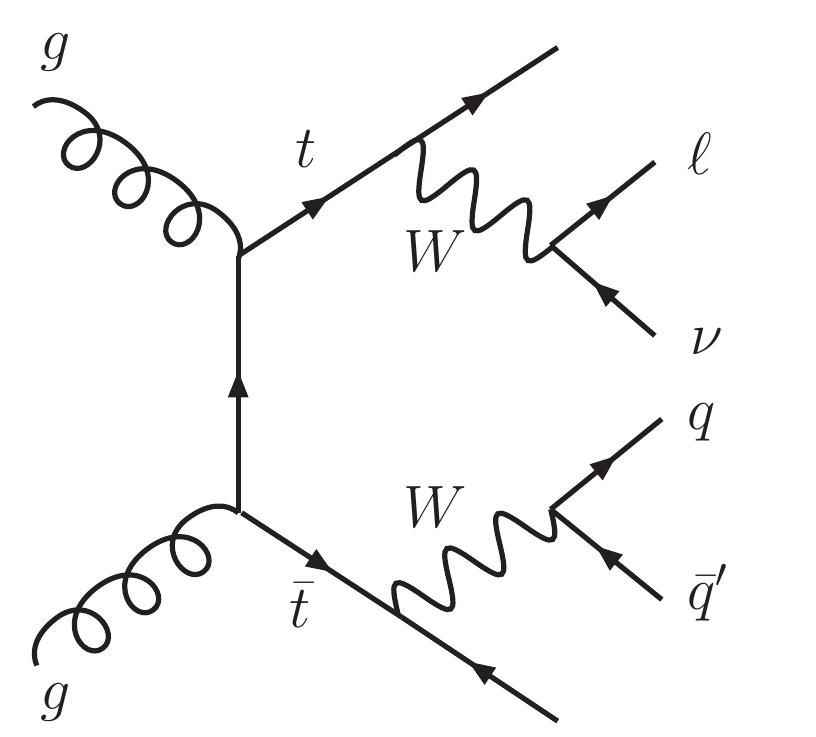}
\includegraphics*[width=0.4\textwidth,height=3.2cm]{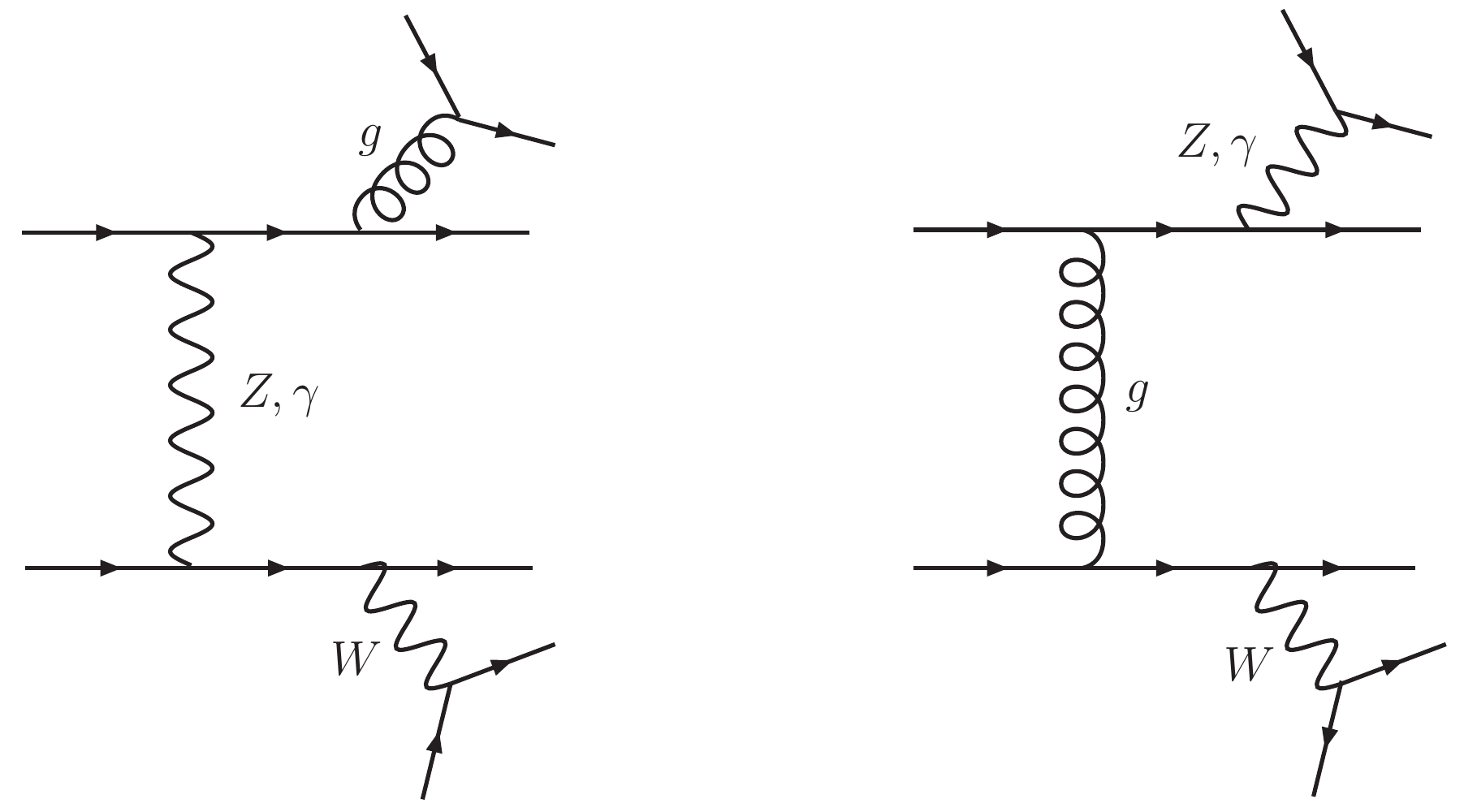}
\begin{picture}(0,0) (0,0)
  \put(-324,-5) {\small{(a)}}
  \put(-224,-5) {\small{(b)}}
  \put(-129,-5) {\small{(c)}}
  \put(-43,-5) {\small{(d)}}
\end{picture}
\caption{Examples of contributions to the QCD irreducible background: $t\bar{t}$
production (a,b) and \VVjj (c,d)}
\label{fig:VBSbckgr_QCD}
\end{figure}
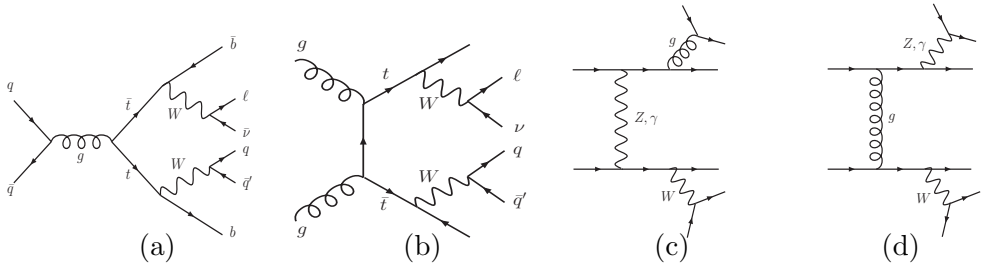

Let us explain with one example which are the processes to be included in a practical case.
Suppose one wants to study  a Higgs produced in boson boson fusion
decaying to two $W$'s with one of the $W$'s decaying leptonically. 
In such a case one will experimentally 
search for a signal of four jets plus a lepton and missing transverse momentum.
The corresponding physical processes that will have to be generated with \Phantom
are all those with a lepton and a neutrino and 4 quarks in the final state at \ordEW +
\ordQCD as well as those with one or two outgoing gluons and three or two quarks in addition
to the leptons. The number of these processes amounts to a few hundreds.
The unweighted generation of all of them will be performed simultaneously with \Phantom.

Analogous considerations hold of course for top antitop or boson boson scattering studies.
One cannot separate the different physical processes just considering the final states,
and the strategy is that of a full calculation. With a complete sample one will thereafter
evidentiate the physics of interest  with appropriate cuts.

\section{Program Description}
\label{sec:descript}

We give in this section a description of the main features of the \Phantom Monte Carlo.
The routine list and an example of input file for running the program are presented
in  appendices A and B respectively. A User Guide can be found in the \Phantom
distribution, with more details on how to run the program and how to prepare the input files. 

\subsection{General features}
\label{subsec:gen_feat}

As already said, \Phantom can generate unweighted event samples for any set of reactions 
with six partons in the final state at $pp$,
$p\overline{p}$ and $e^+e^-$ colliders at \ordEW and \ordQCD.
Possible interferences between the two sets of diagrams can be taken into
account. Alternatively, the user can simulate purely electroweak and \ordQCD
processes separately.
All finite width effects are taken into account as well as all correlations
between the decay of any unstable particle which appear in intermediate states.
In most cases, with the notable exception of \toptop production, the purely
electroweak contribution contains the physical processes we wish to
investigate, while the contributions with one virtual or two external gluons
represent a background.
In particular \ordQCD processes completely
describe ${\mathrm V}{\mathrm V}jj$ productions at hadron colliders and
${\mathrm V}4j$ at $e^+e^-$ machines. 

The program proceeds in two steps which are performed separately. In the first
phase each of the reactions which need to appear in the final event sample is
integrated over and the corresponding cross section is calculated.
At the same time
the density of points in the unit cube employed for the sampling
is recursively modified in order to match the actual behaviour
of the integrand and the integrand maximum is searched for.
The information gathered in this step is stored in the form of
an integration grid.
In the final stage all the grids and maxima are used to produce the unweighted 
event sample.
The separation between the two phases allows for additional flexibility: the
grids produced with a specified set of acceptance cuts can be used to produce
event samples with any set of more restrictive selection cuts or for a smaller
number of reactions without repeating the integration step.
Additional processes can easily be introduced.
  
The number of reactions which contribute to a given final state, say for
instance $4jl\nu$,
can be very large, particularly at hadron colliders.
In our approach it is crucial to keep the number of reactions for which the
individual cross section needs to be computed to a minimum.
By making use of symmetries, the task of computing all relevant processes
can be substantially simplified.
Indeed all reactions which differ by  
charge conjugation and by the symmetry between the first and 
second quark families can be described by the same matrix element, 
provided one take a diagonal CKM matrix and ignores the light quark masses.
This extends
to processes which are related by a permutation of the lepton families.
At most they differ by the PDF's convolution and possibly by a parity
transformation, which can be accounted for automatically. 
In the integration step the user can sum over all processes which are related
by charge and family symmetries setting the flag {\tt i\_ccfam}.

\subsection{Amplitudes}
\label{subsec:amplitudes}

The matrix elements are computed with the
helicity formalism of Refs.\cite{method} and the semi--automatic method
described in Ref.\cite{phact}. In our experience the resulting amplitudes are
much faster than those obtained by completely automatic procedures.
Even though we have not made a detailed comparison with all programs and for
all kind of amplitudes, we happened for instance to find differences
in CPU time of one order of magnitude or more with
{\tt Madgraph}, in agreement with the statements in
Ref.~\cite{JagerOleariZeppenfeld}. 

In our calculation we first of all take advantage of the well known fact that
all processes which share 
the same total particle content, with all partons taken to be outgoing, 
can be described by a single master amplitude.
 
The calculation can be further simplified examining more closely the full set
of Feynman diagrams. Since we assume a diagonal CKM matrix, the
fermion--antifermion pair appearing at the end of a given  fermion line either
are of the same flavour, if only neutral vector insertion or an even number of
$W$ insertions take place along the line, or  reconstruct an isospin doublet
if an odd number of $W$ insertions takes place.
Therefore, limiting ourselves to reactions
in which all external particles are fermions it becomes quickly apparent that
in some processes, fermions can be paired only into neutral
couples (we indicate them as 4Z), e.g. $u\overline{c}\rightarrow u  \overline{c} \mu 
\overline{\mu} e \overline{e}$,
while in others only charged couples can be formed (4W), e.g. 
$u\overline{d}\rightarrow c  \overline{s} \mu 
\overline{\nu}_{\mu} \nu_e \overline{e}$.
In other cases they can form two charged and two neutral
couples (2Z2W), e.g. 
$u\overline{u}\rightarrow b  \overline{b} \mu 
\overline{\nu}_{\mu} \nu_e \overline{e}$.
In all other cases the full set of diagrams splits into a sum of the three basic
gauge invariant sets mentioned above, and can be described  as 
2Z2W+4Z,4W+2Z2W or 4W+2Z2W+4Z, e.g. 
$u\overline{d}\rightarrow u  \overline{d} e 
\overline{\nu}_{e} \nu_e \overline{e}$.
Similarly, all reactions with two external gluons and six fermions can be
computed from two basic amplitudes which, with obvious notation, we call
2g2WZ and 2g3Z.
From the above discussion it becomes clear why in \Phantom we have eight master
amplitudes: three pure electroweak with eight external fermions (4Z, 4W, 2Z2W)  at \ordEW ,
the same three at \ordEW + \ordQCD and the two with two external gluons (2g2WZ and 2g3Z).
We take moreover  advantage of the fact that the diagrams in which two or more of the
external particles are identical can be split in sets which just differ for the exchange
of the identical particles. Therefore the master amplitude corresponds to the basic 
diagrams that one would have if all external particles were different.
In \Phantom the full amplitude of the process at hand is then computed by the
routines
{\tt ampem.f, amp8fqcd.f, amp2g.f } which  call all the necessary master 
amplitudes with the appropriate order of the momenta and call 
{\tt coleval.f} or {\tt colevalew.f} to evaluate the color factors.
The amplitude routines compute separately the sum of the diagrammatic
contributions for every possible color configuration.
The probability used for choosing the color flow of each event, as specified by
the \LHA File Format, is taken to be proportional to the modulus square of the
amplitude of each conÞguration.

The range of reactions which are available in \Phantom is substantially
larger than in \Phase. 
All \ordQCD contributions with one virtual or two external gluons and the 4Z
amplitude were not previously available, only two  master amplitudes
were needed and general routines to compose the different amplitudes and evaluate the color
factors were not necessary. 

\subsection{Master amplitudes computation}

In this section we want to show briefly how the computation of master amplitudes is 
organized. We will first consider a master amplitude for eight external fermions 
at \ordQCD, with one gluon exchange and then the one for 2g2WZ.
We refer to  Ref.\cite{ref:Phase} for a description of the kind of  strategy we use for
\ordEW eight fermion master amplitudes.

\begin{figure}[h!tb]
\centering
\unitlength 1cm
\begin{picture}(15,3) 
\includegraphics*[width=.9\textwidth]{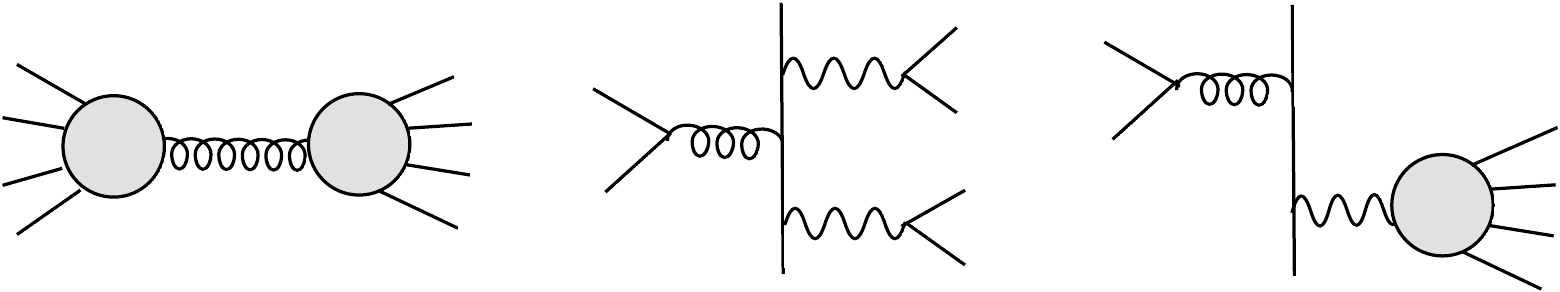}
\end{picture}

\caption{Gluon exchange diagrams}
\label{fig:glexc}
\end{figure}

In the first case the possible ways in which the gluon can be exchanged are depicted in 
\fig{fig:glexc}

\begin{figure}[h!tb]
\centering
\unitlength 1cm
\begin{picture}(15,3) 
\includegraphics*[width=.9\textwidth]{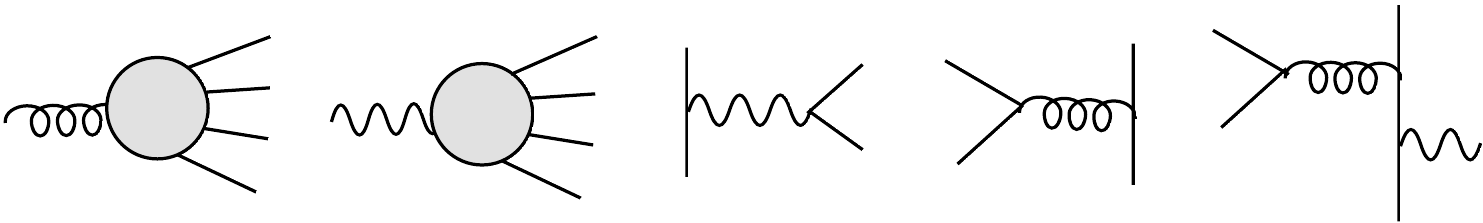}
\end{picture}

\caption{Gluon exchange subdiagrams}
\label{fig:glexcsubdiag}
\end{figure}

This immediately shows that it is convenient to perform the calculation via the
subprocesses  of \fig{fig:glexcsubdiag}. Starting from the left, the first two
correspond to the decay of a virtual boson or gluon to four fermions, the third
and the fourth are building blocks in which two external fermions annihilate in
a virtual boson or gluon which is inserted in the middle of a fermion line.
At least one of the two ends of the fermion line does not correspond to an external
particle.  The other parts  of the fermion line and its attachments can be easily combined
with such  building blocks in the method we use \cite{method,phact}. 

\begin{figure}[h!tb]
\centering
\unitlength 1cm
\begin{picture}(15,3) 
\put(-0.2,0){\epsfig{file=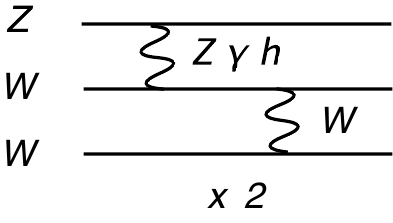,width=.2\textwidth,height=2.5cm}}
\put(3.8,0){\epsfig{file=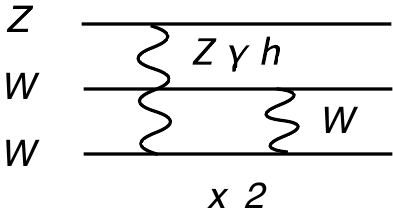,width=.2\textwidth,height=2.5cm}}
\put(7.8,0.5){\epsfig{file=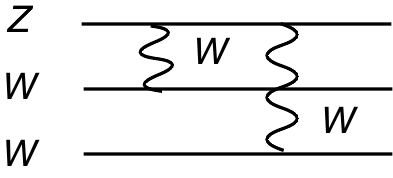,width=.2\textwidth,height=2.cm}}
\put(11.5,0.2){\epsfig{file=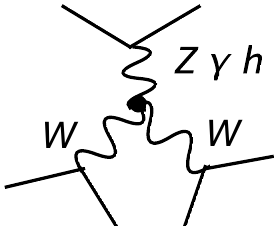,width=.2\textwidth,height=2.5cm}}
\end{picture}

\caption{Electroweak configurations of 2g2WZ}
\label{fig:ewconfig}
\end{figure}

For the master amplitude of 2g2WZ we notice first of all that the possible electroweak 
configurations are those of \fig{fig:ewconfig}. To obtain all possible diagrams
of the master amplitude it is sufficient to attach in all possible ways the two external
gluons. 

\begin{figure}[h!tb]
\centering
\unitlength 1cm
\begin{picture}(15,3) 
\includegraphics*[width=.9\textwidth]{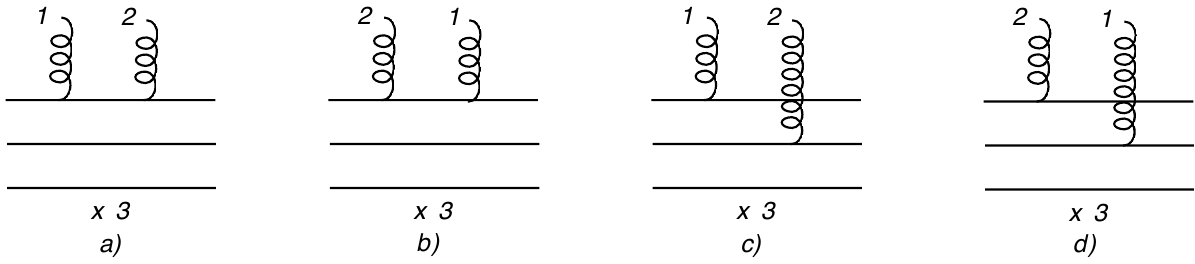}
\end{picture}

\caption{Color structures of 2g2WZ}
\label{fig:colstruct}
\end{figure}

This leads to the twelve color structures of \fig{fig:colstruct} where
the first two starting from left correspond to both gluons attached to the same quark
line in all possible ways and the second two to the two gluons attached to two different  
quark lines. It is superfluous to remind that the configuration with the two external 
gluons forming a triple vertex with a gluon propagator attached to the quark line
can always be reduced to a linear combination of configurations {\it a)} and {\it b)}.

Also for 2g2WZ we proceed evaluating subprocesses in order to avoid computing
twice a subdiagram appearing in several different feynman diagrams. Therefore
one starts with subdiagrams of increasing complexity: virtual boson decaying to
two fermions, to two fermions and one gluon and to two fermions and two gluons.
Then those corresponding to one virtual fermion decaying to three fermions or three
fermions + 1 gluon, etc. Putting together all these pieces one finally arrives at computing
all groups of Feynman diagrams belonging to the the same color structure, which is
precisely the purpose of the routines computing the master amplitudes.

\subsection{Parameters and Parton Distribution Functions}
\label{subsec:parameters}

Standard model parameters are defined in the routine {\tt coupling.f}.
In our notation, {\tt rmw}, {\tt rmz}, {\tt rmt}, and {\tt rmb} are the 
$\PW$, $\PZ$, top and bottom 
masses respectively. The total $\PW$ and $\PZ$ widths are given by
{\tt gamw} and {\tt gamz}. Higgs and top widths are computed in the same
routine by standard formulas. 
\par\noindent
\Phantom employs the $G_\mu$-scheme
defined by the input set: $M_\PW$, $M_\PZ$ and $G_F$. According to this 
scheme, the calculated parameters are   
\beanon
\sin^2\theta_\PW = 1-(M_\PW/M_\PZ )^2~~~~~~~~
\alpha_{em}(M_\PW )={\sqrt{2}\over{\pi}}G_FM^2_\PW \sin^2\theta_\PW
\eeanon
\noindent
where $\theta_\PW$ is the weak mixing angle, and $\alpha_{em}$ the 
electromagnetic fine structure constant. 

We introduce the 
decay width in the propagators of unstable particles using
the fixed-width scheme. 
For the vector boson propagators, in the unitary gauge we work in,
it consists in replacing $M^2$ with 
$M^2-iM\Gamma$ both in the denominator and in the $p^\mu p^\nu$ term.
This scheme preserves U(1) gauge invariance at the 
price of introducing unphysical widths for space-like vector bosons.
A general discussion of the fixed-width scheme as well as of alternative
approaches to introducing decay widths in scattering amplitudes can be found in
the papers in Ref.~\cite{fl}.

For the PDF's we use the Les Houches Accord distribution \cite{LHAPDF}, 
which can be downloaded from http://projects.hepforge.org/lhapdf/ .

\subsection{Iterative-Adaptive multichannel}
\label{subsec:adapt_multich}
For completeness we describe in this section the iterative and adaptive
multichannel technique which \Phantom employs for
integration. In our opinion the ability to adapt is the overriding 
consideration for multidimensional integrals of discontinuous and sharply 
peaked functions. 
An individual process can contain hundreds of diagrams and several unstable
massive particles can appear at intermediate stages. The resonant peaking 
structure of the amplitude is therefore generally very rich. As a consequence 
the 16-dimensional phase space has multiple combinations of 
non-trivial kinematical regions corresponding to enhancements of the matrix
element which need to be smoothed out in order to achieve a good convergence of
the integration process, which in turn results in good efficiency of event
generation.

We have merged
the {\it multichannel} method \cite{multichannel} and the {\it adaptive} 
approach \`a la {\tt VEGAS} \cite{vegas}. 
An algorithm based on a similar 
philosophy has been proposed in \cite{vamp}.
In the {\it multichannel} 
approach, mappings into phase-space variables are chosen in such a way that 
the corresponding Jacobians cancel the peaks of the differential cross section. 
These mappings are not in general unique. One normally needs several different 
phase-space parametrizations, called channels, one for each possible peaking 
structure of the amplitude.
This gives rise to a huge combinatorics which requires a 
correspondingly large number of channels. 
Number and type of these mappings must be fixed a priori, before starting the 
integration. The {\it multichannel} method thus requires a guess on the 
behavior of 
the integrand function. It indeed relies on the expectation that the selected
set of channels, properly weighted \cite{weights}, is able to describe 
reasonably well the amplitude. As no adaptivity is provided (apart from 
the freedom to vary the relative weight of the different channels), neglecting 
even one channel might worsen considerably the convergence of the integral. 

The criteria of {\it adaptive} integration as performed by {\tt VEGAS} are 
rather different. This approach bases its strenght on the ability to deal 
automatically with 
totally unknown integrands. By employing an iterative method, it acquires 
knowledge about the integrand during integration, and adapts consequently 
its phase-space grid in order to concentrate the function
evaluations in those regions where it peaks more. 
{\tt VEGAS} divides the N-dimensional space in hypercubes, and scans the 
integrand along the axes. For a good convergence of the integral, it thus 
requires amplitude peaks to be aligned with the axes themselves. The problem 
can be easily solved if one set of phase-space variables is sufficient 
to describe the full amplitude peaking structure. In this case, the alignement 
can always be obtained by an appropriate variable transformation. The method 
becomes inefficient when it is impossible to align all enhancements with a 
single transformation.

In order to reduce the number of separate channels 
one has to consider in the {\it multichannel}, we use {\it multi-mapping}. 
In general multiple peaks can appear in the same variable, together with 
long non-resonant tails which extend far away from the peaks. This latter case 
is more and more severe as the collider energy increases. For instance
the mass of a neutral fermion pair $f\bar{f}$ would typically 
resonate at both the $Z$ and Higgs mass. The two peaks and the three non resonant 
regions can be separately mapped into a five zone partition of the basic interval [0,1]
of the integration variable.
While {\it multi-mapping} is extremely useful to improve the convergence of 
{\tt VEGAS} integration within a single phase-space parametrization, in
general several such parametrizations are needed. In this case, one has to
introduce $N$ different channels (in standard multichannel language) with 
their proper {\it multi-mapping}. Each channel defines a non-uniform 
probability density $g_i(\Phi )$, which describes a specific class of 
amplitude peaks.
One can then write
\bea
I = \int d\Phi f(\Phi ) & = & \sum_{i=1}^N\alpha_i\int {d\Phi~g_i(\Phi)f(\Phi )
\over{\sum_{i=1}^N\alpha_i~g_i(\Phi )}} = \nonumber \\ 
& & \sum_{i=1}^N\alpha_i\int 
dx_i~f^{\prime}(G^{-1}_i(x_i)) = \sum_{i=1}^N\alpha_iI_i
\label{multigen}
\eea
\par\noindent
$\alpha_i$ being the so called {\it weight} of the {\it i}-th channel 
($x_i=G_i(\Phi )$), and $f^{\prime}(\Phi )$ the smoothed integrand. The 
$\alpha_i$ quantify the relevance of the different peaking structures of the 
amplitude. 
Owing to the very poor knowledge of the integrand, it is rather difficult to 
guess these values a priori. Usually, they are computed and optimized during 
the integration run. 
In the {\it iterative-adaptive multichannel}, the integral in 
\eqn{multigen} splits in $N$ distinct contributions. 
The presence of identical final-state particles increases the possible list of
resonant structures. In order to keep the number of separate integration runs 
manageable, we include all jacobians generated by particle exchange in the
denominator of \eqn{multigen}, while exploiting the freedom to relabel
the momenta to regroup all integration runs related by particle exchange to a 
single one. 

The integration proceeds through two steps: the first one which is
called thermalization collects preliminary information about the integrand and 
fixes the weight of the different channels.
In every thermalizing iteration, all channels are independently integrated 
for some set of $\alpha_i$. At the end of each iteration, a new set of 
phase-space grids (one for each channel), and an improved set of $\alpha_i$ 
are computed. The criteria for weight optimization we adopt is 
\be
\alpha_i={I_i\over{\sum_{i=1}^N I_i}}
\ee
\noindent
where $I_i$ is the $i$-channel integral. The new set of $\alpha_i$ and grids 
are then used in the next iteration. The procedure is repeated until a good 
stability of the $\alpha_i$ is reached. 
Weights smaller than $1\times 10^{-3}$ after the third thermalization iteration
are set to zero and the corresponding channels are discarded. If any number of
channels is dropped two additional thermalization iterations are performed.
In the second step no further change of the weights is allowed, the total
integral for each channel is evaluated an a corresponding phase space grid is
produced. These grids are then  stored and made available for event
generation.

\subsection{Phase space}
\label{subsec:phspace}

While the number of possible combinations of resonances is very
large, from a topological point of view  the structure is much simpler.
Apart from the weak vector bosons the only unstable particles in the SM
which are
relevant for LHC or ILC physics are the Higgs boson and the top quark.
While the weak
bosons at tree level have only two body decays, the top quark decays to $bW$ and
then to a three body system. For the Higgs boson the situation is more complex
and the ratio of the various branching ratios and therefore their
phenomenological relevance depends crucially on the Higgs
mass. The Higgs decay channels however fall into two groups: two body decays
like $\gamma\gamma$, $b\overline{b}$ or $\tau^+\tau^-$ or four body decays
through $W^+W^-$ or $ZZ$.
It is therefore possible to construct a limited number of phase space
parametrizations each of which,
varying the masses and widths of the particles which appear in the intermediate
states, can describe several channels in the multichannel sense.

In addition to $s$--channel resonances,
$t$--channel enhancement can also be present. At the LHC they are regulated by
minimum $p_T$ requirements. At the ILC the issue is much more relevant because
processes with electrons lost in the beam pipe can be an important background
to charged current reactions.

Since we do not keep track of individual diagrams, the possible resonant
structures must be deduced from the external particle content only.
This is performed in the {\tt proc.f} routine.
Notice that the resonant structure depends on the perturbative order:
$qq \rightarrow qqWW$ can have a $H\rightarrow WW$ resonance 
at \ordEW but not at \ordQCD.

The criteria we adopt to automatically define 
number and type of channels needed for a given process are the following:
for each process we consider all particles as outgoing and
then examine all sets of four
$f_i\bar{f}_j$ pairs which can be constructed, retaining only those in which all
pairs can be identified with the decay products of a $W$ or $Z$ boson.
This produces the complete set of
possible two--particle enhancements. As a second step, for each 
selected set of pairs we search for triplets which reconstruct $bW^+$ and
$\bar{b}W^-$ states and $W^+W^-$ and $ZZ$ states respectively. Pairs in which
one fermion is outgoing and one incoming are associated with $t$-channel
enhancements.
For each set we determine the maximum number of enhancements which can be 
simultaneously present and introduce the corresponding channel in the
list of channels utilized for the process.
Multi-mapping and adaptivity will take care of all related partially-resonant 
or non-resonant configurations.
Each channel in \eqn{multigen} is integrated separately with {\tt VEGAS}.

\begin{figure}[h!tb]
\centering
\unitlength 1cm
\begin{picture}(15,10.5) 
\put(0,7.5){\epsfig{file=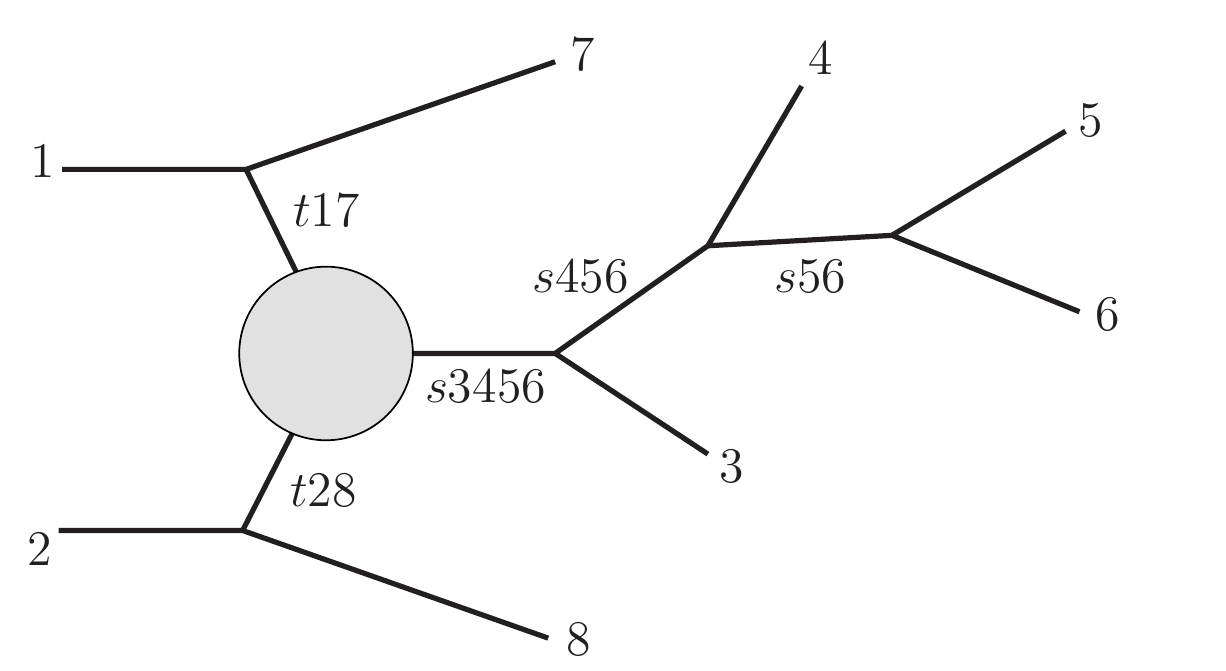,width=0.30\textwidth}}
\put(5,7.5){\epsfig{file=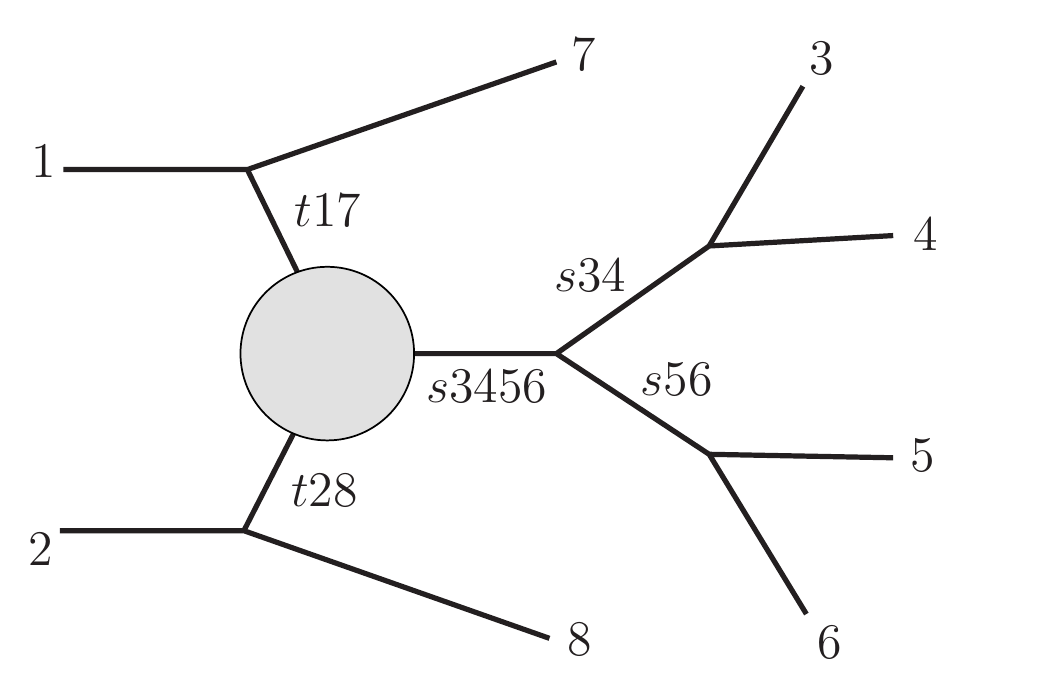,width=0.30\textwidth}}
\put(10,7.5){\epsfig{file=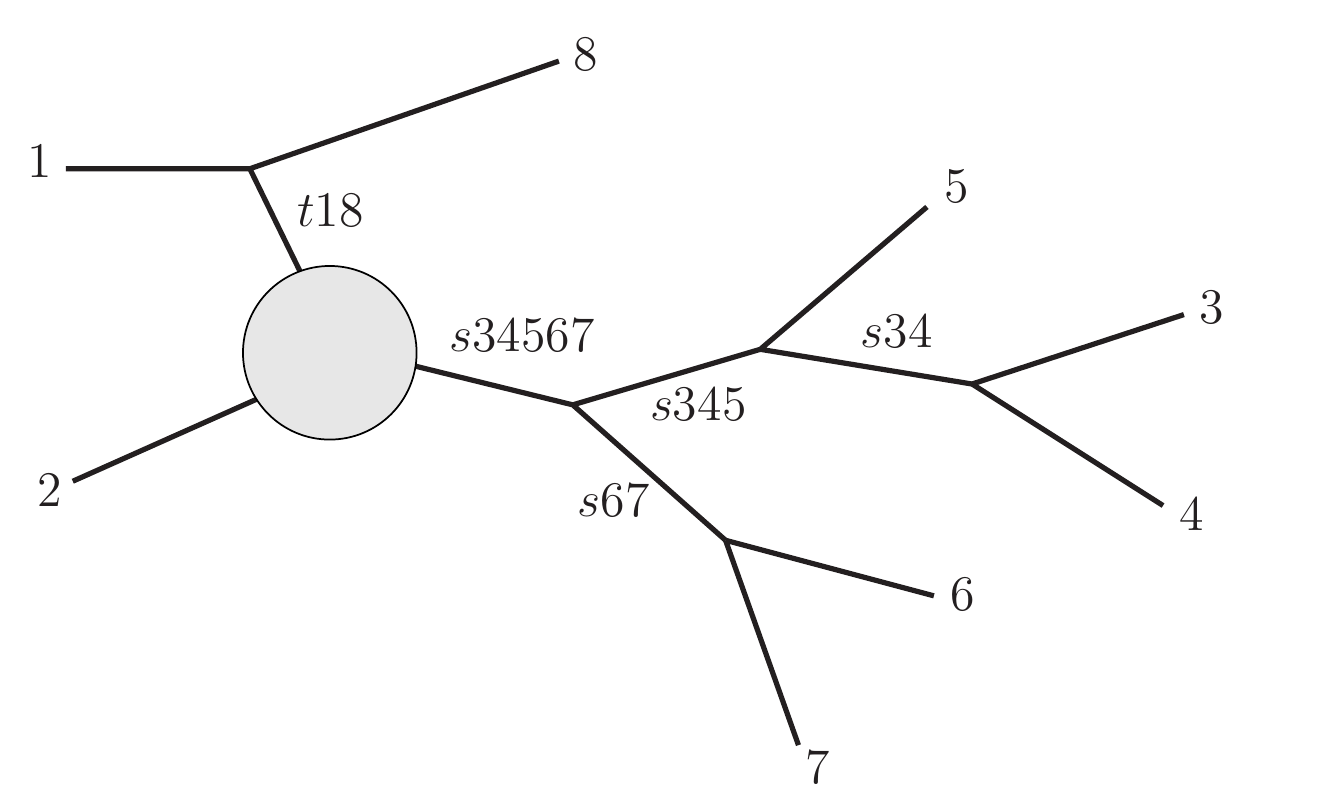,width=0.30\textwidth}}
\put(0,4.3){\epsfig{file=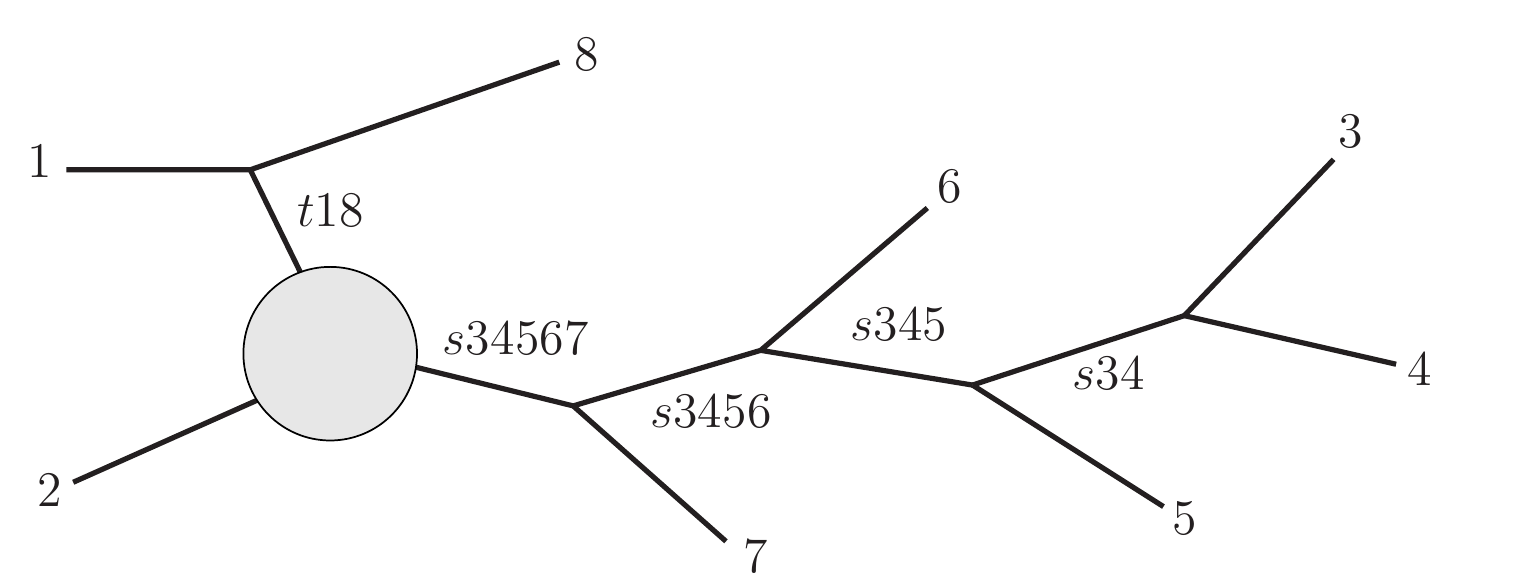,width=0.30\textwidth}}
\put(5,3.8){\epsfig{file=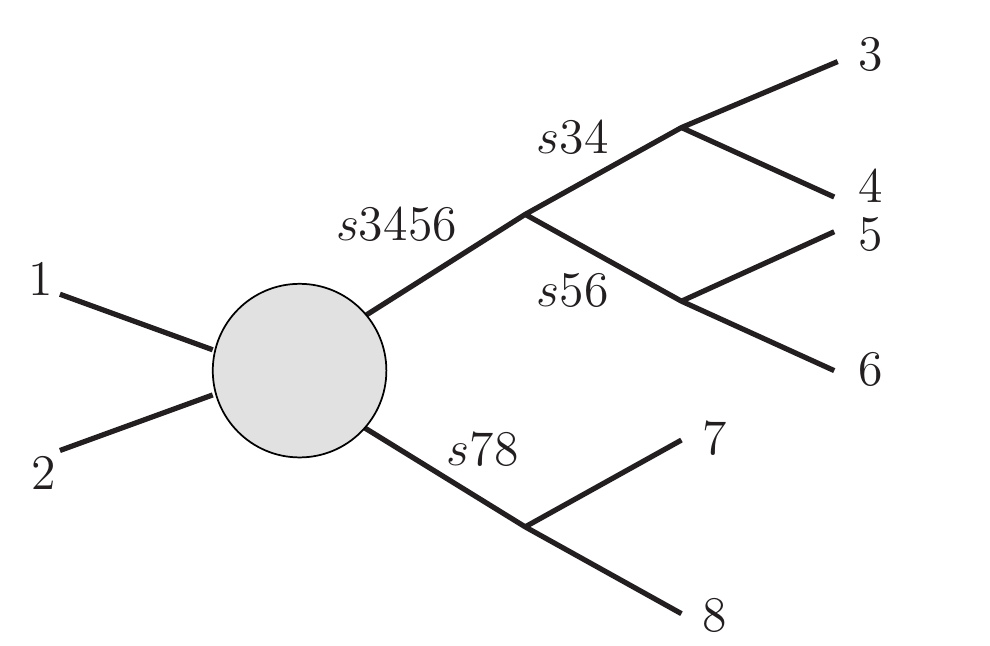,width=0.30\textwidth}}
\put(10,3.8){\epsfig{file=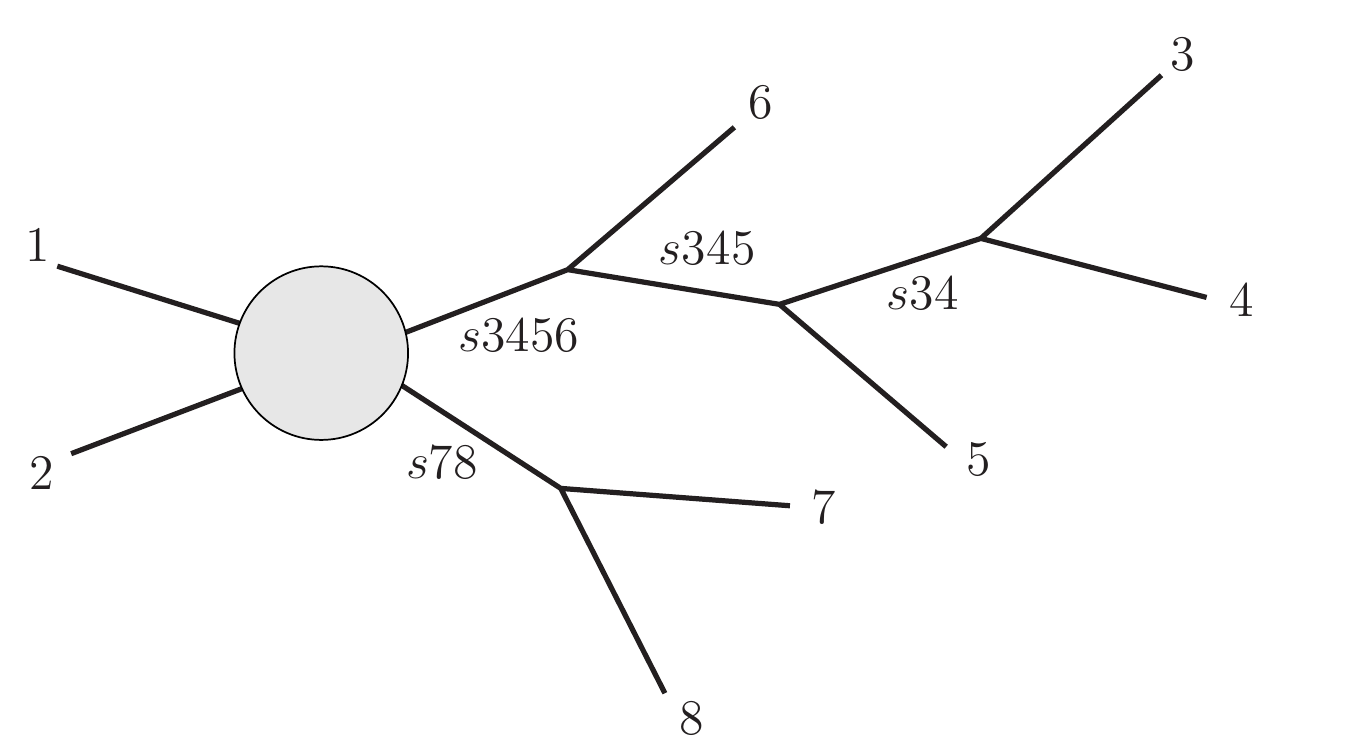,width=0.30\textwidth}}
\put(5,0){\epsfig{file=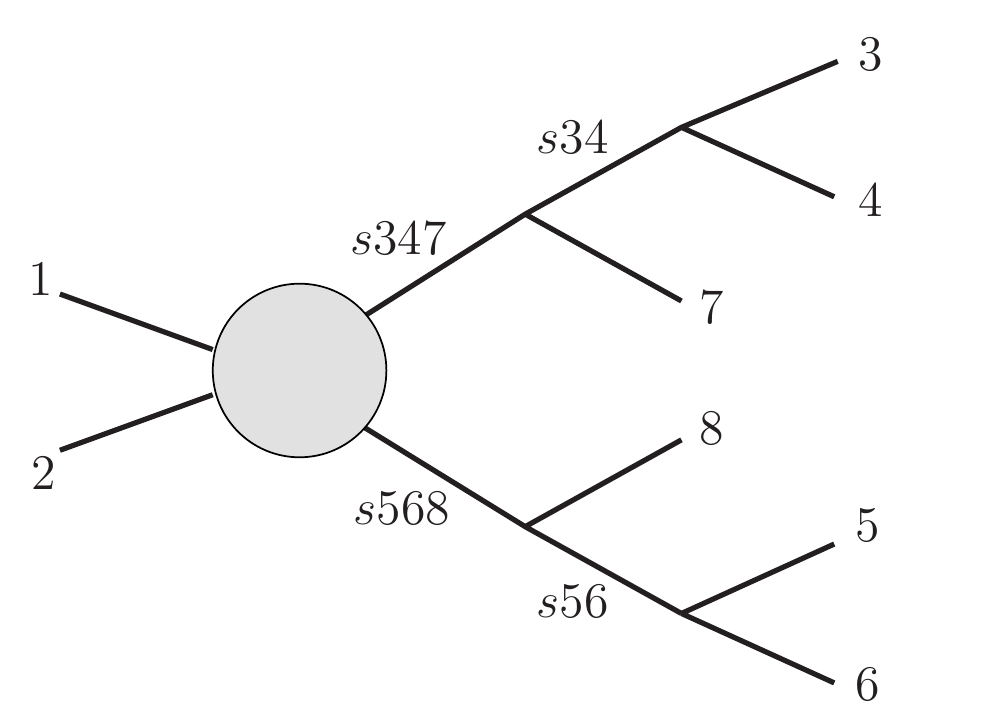,width=0.30\textwidth}}
\end{picture}

\caption{Examples of phase space mappings}
\label{fig:phsp1}
\end{figure}

In \fig{fig:phsp1} we show pictorially  the flexible
phase space mappings which
are available in \Phantom . We denote by $\Omega_{ij}$ the usual
angular variables in the center of mass of particles $i$ and $j$
and define $s_{i_1\dots i_n} = (p_{i_1}+\dots +p_{i_n})^2$ and
$t_{ij} = (p_i-p_j)^2$.
Then, for instance, the mapping
corresponding to the central figure of the first row will use as variables $t_{17}$, $t_{28}$,
$s_{3456}$, $\Omega_{(34)(56)}$, $s_{34}$, $\Omega_{34}$, $s_{56}$, 
$\Omega_{56}$,
$s_{34567}$. The remaining variables are the  azimuthal angles $\phi_8$ and
$\phi_7$ of particle 8 in the overall centre of mass and of particle 7 in the
center of mass of 34567, respectively.
If we indicate with  $M_{X}$ the invariant mass at which the $s_{X}$ variable
can resonate or the mass of the vector boson which enters into the $t_{X}$
channel exchange, we could have $M_{17}=M_{28}=M_{34}=M_{56}=M_W$,
$M_{3456}=M_H,M_Z$ which describes a $WW$ fusion channel with a Higgs resonance;
on the other hand we could also have $M_{17}=0$ ,$M_{28}=M_{56}=M_W$, 
$M_{34}=M_Z$ which describes a $\gamma W \rightarrow ZW$ virtual scattering.
It is clear that a large number of possibile channels can be described by this
single parametrization.

\section{Running the program}

\subsection{Modes of operation}
\label{subsec:modes}
The program has two modes of operation which are selected by the input value 
{\tt ionesh}. If {\tt ionesh=0} the program computes the cross section for
one single process, specified by {\tt iproc}. If the input variable {\tt iflat}
is set to {\tt 1} the program produces one or more integration grids
depending on the number of channels required for a good mapping of phase space
for the selected process.  
If {\tt ionesh=1} the program generates unweighted events for a set of
processes, which is specified by the user, using the previously produced grids.
All the integration grids for each selected process must be included for a
meaningful generation.

When {\tt ionesh=0} the program proceeds in two steps. The first one is called
thermalization.
It determines the relative weight of each channel in the multichannel 
integration and it
produces a first instance of \Phantom space grids, one per channel.
At least three thermalization iterations are performed. At the end of the third
iteration all channel whose relative weight is smaller than $10^{-3}$ are
eliminated. If any channel is discarded, two additional thermalization iterations
are performed, regardless of the flag {\tt itmx\_therm} mentioned below.
The grids produced in the thermalization stage
are then used as a starting point for the second step which consists
of one integration per channel. Each integration will typically consist of
several iterations and at each iteration the \Phantom space grid will be refined
in an effort to decrease the overall variance. 
A number of iterations between 3 and 5 is normally the best choice.
If higher precision is requested it is usually more convenient to 
increase {\tt ncall} rather than {\tt itmx}.
The user must be aware of the fact that if no point survives the cuts during
an iteration, either during thermalization or at the integration stage, \veg
will stop with an error.

At $pp$ and $e^+e^-$ colliders, when {\tt ionesh=0} the process is computed
exactly as specified by the user,
assuming the first incoming particle to be moving in the 
$+z$ direction and the second one in the $-z$ direction.
Even though at a $pp$ collider this violates the symmetry between the two
initial state protons, it can be useful for testing and for specialized studies.
The full cross section can be easily obtained for unlike incoming partons by
symmetrizing all distributions with respect to the beams and multiplying by a
factor of two.
At a  $p\bar{p}$ collider the default for unlike incoming partons is to sum over
the two possible assignement of the two partons to the beams.
The behaviour at $pp$ colliders can, and typically should,
be modified when {\tt ionesh=1} by the flag
{\tt i\_exchincoming}.

 \subsection{Input}
\label{subsec:input}
The syntax of the input is almost identical to the one required by the
CERN library routine {\tt FFREAD}. Routines internal to \Phantom are however 
used ({\tt iread, rread}),
so that real variables can (and must) be given in double precision.

All lines in the file of input must not exceed 80 characters, with the exception
of filenames which can be 200 characters long.
A {\tt \large *} or {\tt \large C} character at the beginning of a line
identifies it
as a comment line. Comment lines can be freely interspersed within the input
file,
with the only obvious exception that they must not interrupt a list
of input values for a single array variable.
The name of the variable to be read must be specified as
the first word of a line 
(needs not to begin in column 1). Its value(s) must follow it. 
The list of values can span several lines.
Variables which are  not needed for the process under study 
will be ignored. They can be left in the file without harm.
All variables actually read from the input file will
be reproduced in the output.

The input variable { \tt iproc} specifies the desired process using the
standard Monte Carlo particle numbering scheme:
 
\begin{table}[hbt]\centering
\begin{tabular}{|c|c|c|c|c|c|c|c|c|c|c|c|}
\hline
 $d$ & $u$ & $s$ & $c$ & $b$ & $e^-$ & $\nu_e$ & $\mu$ & $\nu_\mu$ &
  $\tau$ & $\nu_\tau$ & $g$\\
\hline
 $1$ & $2$ & $3$ & $4$ & $5$ & $11$ & $12$ & $13$ & $14$ &
  $15$ & $16$ & $21$\\
\hline
\end {tabular}
\end {table}
\noindent
Antiparticle are coded with the opposite sign.
The first two entries represent the initial state partons. Therefore the
string
\be
3 \quad  -4 \quad \;   2 \quad  -2 \quad \;  3 \quad  -3 \quad \; 13 \quad  -14
\ee
corresponds to the reaction
\be
s\bar{c} \rightarrow u \bar{u} s \bar{s}  \mu \overline{\nu_\mu}.
\ee

\subsection{Cuts}
\label{subsec:cuts}

The program provides two means of specifying acceptance cuts. A basic predefined
set can be imposed through the input file. They cover the usual requirement of
minimum energy, transverse momentum and separation among the final state
partons.
All cuts are specified by an integer of the type
{\tt i\_flag} which specifies whether the corresponding
cut is activated ({\tt i\_flag=1}) or not ({\tt i\_flag=0}) 
and by one or more values which define the extrema of the
accepted region. The name of the flag in most cases is the name of the
corresponding variable with  {\tt i\_} prepended. 

The first part of the input cuts is common to 
both running modes, and must be always kept unchanged when passing from 
{\tt ionesh=0} to {\tt ionesh=1}. It constitutes in fact the setup under which 
phase-space grids are produced. In order to give the possibility of imposing 
other cuts at generation level, the {\it input-file} has also a cut section 
specific of the {\it one-shot} mode,
The corresponding variables 
are the same as those in the common input section; they are just renamed with 
a suffix  - {\tt os} - appended. These additional cuts, operating at 
generation level, are obviously effective only if more restrictive than the 
common ones.
Since no predefined set of cuts will be able to cover all possible user needs,
it is also possible to include 
extra user-specified cuts via a routine called {\tt iuserfunc}, 
an example of which is provided in the program package, inside the file {\tt cuts.f} 
As for the predifined set of cuts, additional requirements can be imposed at
generation level via the {\tt iuserfuncos} routine.
If user-specified cuts are imposed the package needs to be recompiled.

\subsection{Output}
\label{subsec:output}

The output of the program depends obviously on the mode of operation. The result
of running the program in the integration phase ({\tt ionesh=0}) 
is an output file in which are reported the input choices and the result of the various
iterations during  integration for the different channels. One can check from
this file the accuracy reached by the integration and the reliability of the result.
The informations needed for the successive unweighted event generations are given for
 channel {\sl xx} of a given process by files under the name {\tt phavegas{\sl xx}.dat}.
For every process a separate directory must be created.
The output of the  ({\tt ionesh=1}) mode is a file in which the statistics for each
channel contributing to the generation are reported, as well as the total integral
of all channels contributing to the generations. It is advisable
to check that this result
corresponds, within the statistical errors, to the sum of the cross sections of the
various processes contributing to the final sample, taking into account possible
factors of two due to the  exchange of incoming particles. If one has set in input  
{\tt iwrite\_event 1}, as it is normally the case, the unweighted events generated 
at parton level are written in a file named {\tt phamom.dat}
using the \LHA File Format \cite{LHAFF}. The way the events are written depends 
on the flag {\tt iwrite\_mothers}. If this is set to 1  additional information 
is added to the event concerning possible ``mothers'' (resonances) which  
generate final particles or intermediate ``on shell'' particles. This information
is used by the hadronization Monte Carlo's and in particular by \Pythia  \cite{PYTHIA6.4}
for the evolution of the parton shower and the hadronization.  
Preliminary studies show that the use of this information has sizeable consequences, at 
least for top processes, on the  final fully hadronized events.\footnote{We 
thank Roberto Chierici for pointing out this feature to us and Torbjorn
Sjostrand and Fabio Maltoni for discussions on this point.} It is obvious that in a
full calculation, the identification of the ``mothers'' in an event
is not straightforward as it would be in a production times decay approximation.
We have therefore chosen to somehow rely on the kinematic configuration of the event for 
assigning the ``mothers''. We first of all determine which are the possible mothers
considering the particular phase space channel and parameters used to produce the
kinematical configuration. We then look at the color configuration chosen via the
amplitudes and test if the two informations agree. Only to  those resonances which
pass the check we assign the role of ``mothers''.

\subsection {Scripts}
\label{subsec:scripts}
Determining the minimal set of reactions for which grids need to be computed
can be tedious and error prone.
Therefore we have included in the distribution two Perl scripts
{\tt setupdir-all-LSF.pl} and {\tt setupdir-all-LSF\_ILC.pl} which handle the
task for hadron and $e^+e^-$ colliders respectively.
The script can be executed with, for instance:
\begin{verbatim}
perl setupdir-all-LSF.pl [-options ...] 
\end{verbatim}
\vsk

where options include:
\begin{small}
\begin{verbatim}
  -basedir      directory which contains the exe file   
                 (full pathname)                       [./]
  -dirtreeroot  root of new tree (full pathname)       [mh0]
  -template     input template file                    [template.st0]
  -executable   executable file                        [phantom.exe]
  -inputstring  must contain only leptons and gluons   [""]
                must be enclosed in double quotes: 
                 "e e_ mu vm_" 
                only processes containing -inputstring
                will be generated 
  -quarks       number of quarks to be inserted        [0]
                 (even integer > 0 and <= 10)    
  -Top          number of intermediate top/antitop quarks which can 
                be reconstructed by the final state particles 
                (possible values 0,1,2). If the option is of the form 
                n+ any reaction with at least n intermediate tops are 
                accepted. If the option is not set any number of tops 
                is accepted (equivalent to 0+). 
  -help         prints usage details                      
\end{verbatim}
\end{small}
Long options can be abbreviated up to one letter, eg.
{\tt -basedir  X} can be passed as {\tt -b X}. In square brackets the default
values are reported.
For each reaction the scripts create a directory which contains the
corresponding {\tt r.in} input file which is generated using the given template which must
contain all input parameters with the exception of {\tt iproc} which is supplied
by the script. A runfile called {\tt run} is also created.
In the  root directory of the new tree the script creates a file called
{\tt LSFfile} which can be used for submitting all
integration jobs to the CERN LSF batch system. For different batch systems the
user should appropriately edit the Perl scripts.

\section{\ordEW\unskip+\ordQCD results in the semileptonic $\mu \nu_\mu$ channel}
\label{sec:example}
In this section we present an example of some phenomenological results
obtained with \Phantom.
These studies have been performed via the generation of high statistic samples of 
unweighted events at parton level for all processes with a pair of leptons 
$\mu\bar\nu_\mu$  and different Higgs masses at the LHC.
For a more complete discussion of other studies performed  with \Phantom  we refer to
\cite{Accomando:2006vj}.   

We investigate how much the sensitivity to the \VV scattering signal 
in the semileptonic $\mu \bar\nu_\mu$ channel
is affected by the QCD irreducible background.

In the absence of firm predictions in the strong scattering
regime, trying to gauge the possibilities of discovering signals of
new physics at the LHC requires the somewhat arbitrary definition of a model of
\VVL scattering beyond the boundaries of the SM. Some of these models predict
the formation of spectacular resonances which will be easily detected.
For some other set of parameters in the models only rather small effects are
expected \cite{unitarization,butterworth02}.

Two groups \cite{Giudice:2007fh,Barbieri} have recently
tried to parametrize the low
energy behaviour of large classes of composite Higgs models. They
showed that the most characteristic signature is a reduced coupling between the
Higgs state and the SM vector bosons, leading to only a partial cancellation of 
the growth of \VVL scattering amplitudes which would eventually result in
violations of unitarity.

The predictions of SM in the presence of a very heavy Higgs provide an upper
bound on the observability of such effects. 
The linear rise of
the cross section with the invariant mass squared \cite{LowEnergyTheorem}
in the hard \VV
scattering will be swamped by the decrease of the parton luminosities at large
momentum fractions and, as a consequence, will be particularly challenging to
detect.  

In our study, we consider the full set of parton-level processes involved at 
\ordEW\unskip+\ordQCD
\begin{eqnarray}
& q q \rightarrow q q q q \mu \bar{\nu}_{\mu} \nonumber  \hspace{0.8cm}
g g \rightarrow q q q q \mu \bar{\nu}_{\mu} \nonumber & \\
& g q \rightarrow g q q q \mu \bar{\nu}_{\mu} \nonumber  \hspace{0.8cm}
q q \rightarrow g g q q \mu \bar{\nu}_{\mu} \nonumber &
\end{eqnarray}
together with a selection procedure as close as possible to the actual 
experimental practice, without resorting to any flavour information other than
the one which a typical $b$-tagging algorithm is able to provide.
A more complete analysis including the $4jW$ background at \ordQCDsq is left for
a forthcoming paper.

The $p_z$ of the neutrino is approximately reconstructed requiring the 
invariant mass of the two leptons to be equal to the $W$ boson nominal mass. Its
transverse momentum is assumed to be equal to the missing $p_T$.

All the results presented in this section have been obtained using the CTEQ5L 
PDF\cite{CTEQ5} set. The QCD coupling constant has been evaluated at the scale

\begin{equation}
\label{eq:PDFScale2}
Q^2 = M_{top}^2 + p_T(top)^2,
\end{equation}
where $p_T(top)$ is the transverse momentum of the reconstructed top, for all
processes in which a $t$ or $\bar{t}$ can be produced. For all  other 
processes the scale has been evaluated as 
\begin{equation}
\label{eq:PDFScale1}
Q^2 = M_W^2 + \frac{1}{6}\,\sum_{i=1}^6 p_{Ti}^2,
\end{equation}
where $p_{Ti}$ denotes the transverse momentum of the $i$-th final state 
particle.

\begin{table}[h!tb]
\begin{center}
\begin{tabular}{|c|c|c|}
\cline{1-3}
 \multicolumn{1}{|c|}{\scriptsize\boldmath$M_H = 200$ \textsl{\textbf{GeV}}} & $\sigma_{\scriptscriptstyle{{EW}}}$
 & $\sigma_{\scriptscriptstyle{EW + QCD}}$ \\
\hline
all events      & 0.89 pb  &  80.8 pb  \\
\hline
top events      & 0.52 pb  &  71.6 pb  \\
\hline
ratio top/all   & 0.58     &  0.89     \\
\hline
\end{tabular}
\end{center}
\caption{Contribution of $t\bar{t}$/single $t$ to the total cross section with 
standard acceptance cuts only (see the left part of Tab.\ref{tab:cuts_munuqcd}).
Comparison between results at \ordEW (EW) and \ordEW\unskip+\ordQCD (EW+QCD).
Interferences between the two perturbative orders are neglected.}
\label{tab:xsec_EW_vs_EW+QCD}
\end{table}

It should be clear from the results shown in Tab.\ref{tab:xsec_EW_vs_EW+QCD} 
that suppressing the top background is the primary goal to achieve. 
In this analysis we assume the possibility to tag $b$-jets in the region
$\vert \eta \vert < 1.5$ with 0.8 efficiency, which allows to discard part 
of the events containing $b$ quarks in the final state.
We impose additional cuts against the top on the invariant mass of triplets of type 
$\{jjj\}$ and $\{j\mu\nu\}$, where $j$ denotes any final-state quark or gluon. 
In order to isolate two vector boson production, kinematical cuts are
applied on the invariant mass of the two most central jets, which are associated
in our analysis to a $W$ or $Z$ decaying hadronically. 
The \VV fusion signature is further isolated by requiring a minimum 
$\Delta \eta$ separation between the two forward/backward jets.

\begin{table}[h!tb]
\begin{center}
\begin{tabular}{|c|c|} 
\cline{1-1} \cline{2-2} 
\textbf{Acceptance cuts} & \textbf{Selection cuts} \\
\cline{1-1} \cline{2-2}
$p_T(\ell^\pm,j) > 10$ GeV & $b$-tagging for $\vert \eta \vert < 1.5$ 
 (80\% efficiency) \\
\cline{1-1} \cline{2-2}
$E(\ell^\pm,j) > 20$ GeV & $\vert M(jjj;j\ell^\pm\nu_{rec}) - M_{top} \vert >15$ GeV\\
\cline{1-1} \cline{2-2}
$\vert \eta(\ell^\pm) \vert < 3$ & $70$ GeV $< M(j_cj_c) < 100$ GeV \\
\cline{1-1} \cline{2-2}
$\vert \eta(j) \vert < 6.5$ & $M(j_fj_b)<70$ GeV ; $M(j_fj_b)>100$ GeV \\
\cline{1-1} \cline{2-2}
$M(jj) > 20$  GeV & $M(jj) > 60$ GeV  \\
\cline{1-1} \cline{2-2}
\multicolumn{1}{c|}{} & $\Delta\eta(j_fj_b) > 4$ \\
\cline{2-2}
 \multicolumn{1}{c|}{} & $ p_T(\ell^\pm\nu_{rec}) > 100$ GeV \\
\cline{2-2}
 \multicolumn{1}{c|}{} & $ \eta(\ell^\pm\nu_{rec}) < 2$ \\
\cline{2-2}
\multicolumn{1}{c|}{} & $M(j_{f/b}\ell^\pm\nu_{rec}) > 250$ GeV \\
\cline{2-2} 
\end{tabular}
\caption{List of kinematical cuts applied in all results on the 
$\mu\nu_{\mu}$ channel. $j$ denotes any final-state quark or gluon, while 
$\ell^\pm$ is the charged lepton. The subfixes $c$,$f$,$b$ mean 
\textit{central}, \textit{forward}, \textit{backward} respectively.
$\nu_{rec}$ is the neutrino reconstructed following the prescription described in the text. 
}
\label{tab:cuts_munuqcd}
\end{center}
\end{table}

\begin{table}[h!tb]
\begin{center}
\begin{tabular}{|c|c|c|c|c|c|} 
\cline{1-5}
\multirow{2}{*}{\footnotesize\boldmath\ordEW} & 
    \multicolumn{2}{|c|}{no Higgs} & \multicolumn{2}{|c|}{$M_H = 200$ GeV} & 
    \multicolumn{1}{c}{}\\
\cline{2-6}
 & $\sigma$ & $\mathcal{L}$=$100\,\mbox{fb}^{-1}$
 & $\sigma$ & $\mathcal{L}$=$100\,\mbox{fb}^{-1}$ & ratio  \\
\hline
all events & 12.46 fb & 1246 $\pm$ 35 & 13.57 fb & 1357 $\pm$ 37 & 0.918 \\
\hline
$M_{cut}=0.8$ TeV & 3.19 fb & 319 $\pm$ 18 & 1.45 fb & 145 $\pm$ 12 & 2.200 \\
\hline
$M_{cut}=1.2$ TeV & 1.28 fb & 128 $\pm$ 11 & 0.41 fb & 41 $\pm$ 6 & 3.122 \\
\hline
$M_{cut}=1.6$ TeV & 0.60 fb & 60 $\pm$ 8 & 0.14 fb & 14 $\pm$ 4 & 4.286 \\
\hline
\end{tabular}
\caption{Integrated \ordEW cross section for 
$M(j_cj_cl\nu)>M_{cut}$ and number of expected events after one year at high 
luminosity having applied the cuts listed in Tab.\ref{tab:cuts_munuqcd}.}
\label{tab:xsec_EW}
\end{center}
\end{table}

\begin{table}[h!tb]
\begin{center}
\begin{tabular}{|c|c|c|c|c|c|} 
\cline{1-5}
\multicolumn{1}{|c|}{\footnotesize\boldmath\ordEW +} & \multicolumn{2}{|c|}{no Higgs} & \multicolumn{2}{|c|}{$M_H = 200$ GeV} & \multicolumn{1}{c}{}\\
\cline{2-6}
\multicolumn{1}{|c|}{\footnotesize\boldmath\ordQCD} & $\sigma$ & $\mathcal{L}$=$100\,\mbox{ fb}^{-1}$
 & $\sigma$ & $\mathcal{L}$=$100\,\mbox{ fb}^{-1}$ & ratio  \\
\hline
all events & 40.70 fb & 4070 $\pm$ 64 & 40.73 fb & 4073 $\pm$ 64 & 0.999 \\
\hline
$M_{cut}=0.8$ TeV & 7.61 fb & 761 $\pm$ 28 & 5.14 fb & 514 $\pm$ 23 & 1.481 \\
\hline
$M_{cut}=1.2$ TeV & 2.53 fb & 253 $\pm$ 16 & 1.73 fb & 173 $\pm$ 13 & 1.462 \\
\hline
$M_{cut}=1.6$ TeV & 1.00 fb & 100 $\pm$ 10 & 0.55 fb & 55 $\pm$ 7 & 1.818 \\
\hline
\end{tabular}
\caption{Integrated \ordEW + \ordQCD cross section for 
$M(j_cj_cl\nu)> M_{cut}$ and number of expected events after one year at high 
luminosity having applied the cuts listed in Tab.\ref{tab:cuts_munuqcd}.
Interferences between the two perturbative orders are neglected.}
\label{tab:xsec_EW+QCD}
\end{center}
\end{table}

At this stage, however, any attempt to appreciate differences between
Higgs and no-Higgs scenarios at high invariant masses would still be vain.
This is essentially due to the fact that the contribution of the 
QCD diagrams is not substantially 
affected by the above-mentioned selection criteria.
Investigating the differences between the kinematics of \VV scattering and 
\VVjj, we have identified additional cuts that serve our purpose.
As the background dominates in the phase space regions characterized by 
vector boson emitted by the tag jets, 
a viable method of taming \VVjj consists in applying cuts on the $p_T$ 
and $\eta$ of the $W$ reconstructed from leptons as well as on the invariant 
mass of the $W$ plus any of the two tag jets.
All details about the selection cuts applied are reported in 
Tab.\ref{tab:cuts_munuqcd}.

\begin{figure}[h!tb]
\centering
\includegraphics*[width=0.49\textwidth,height=6.4cm]{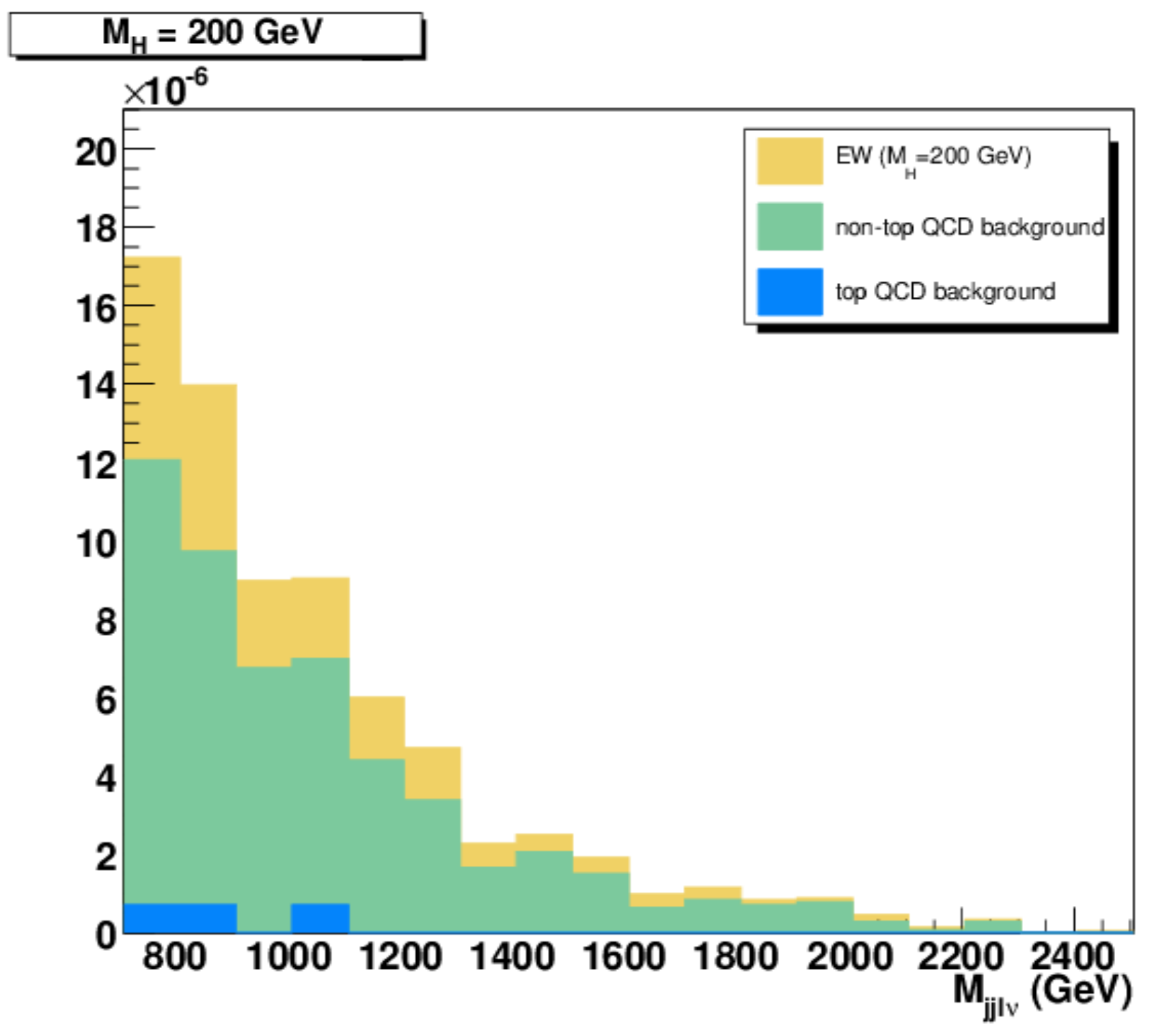}
\includegraphics*[width=0.49\textwidth,height=6.4cm]{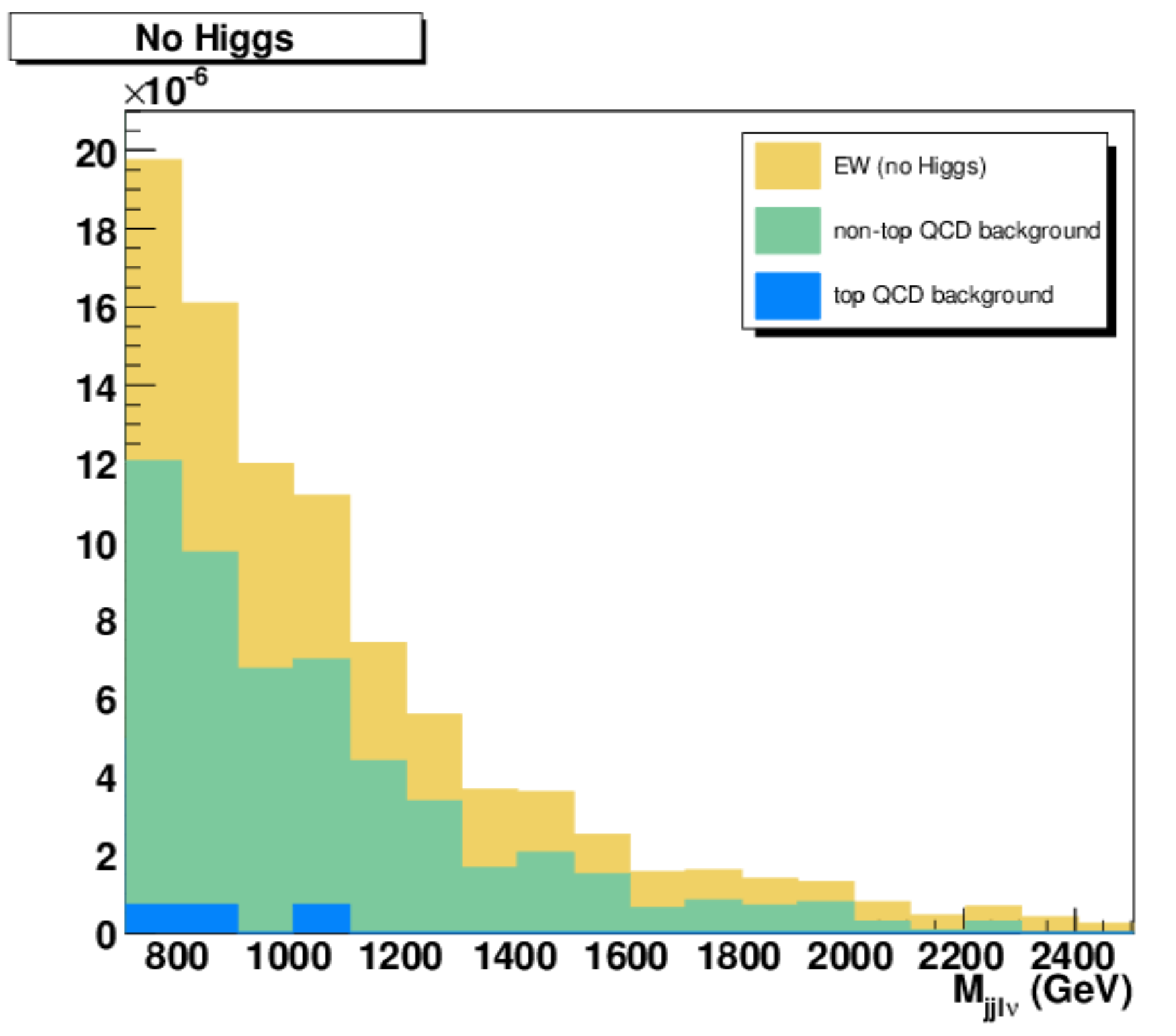} \\
\vspace{0.6cm} 
\includegraphics*[width=0.49\textwidth,height=6.4cm]{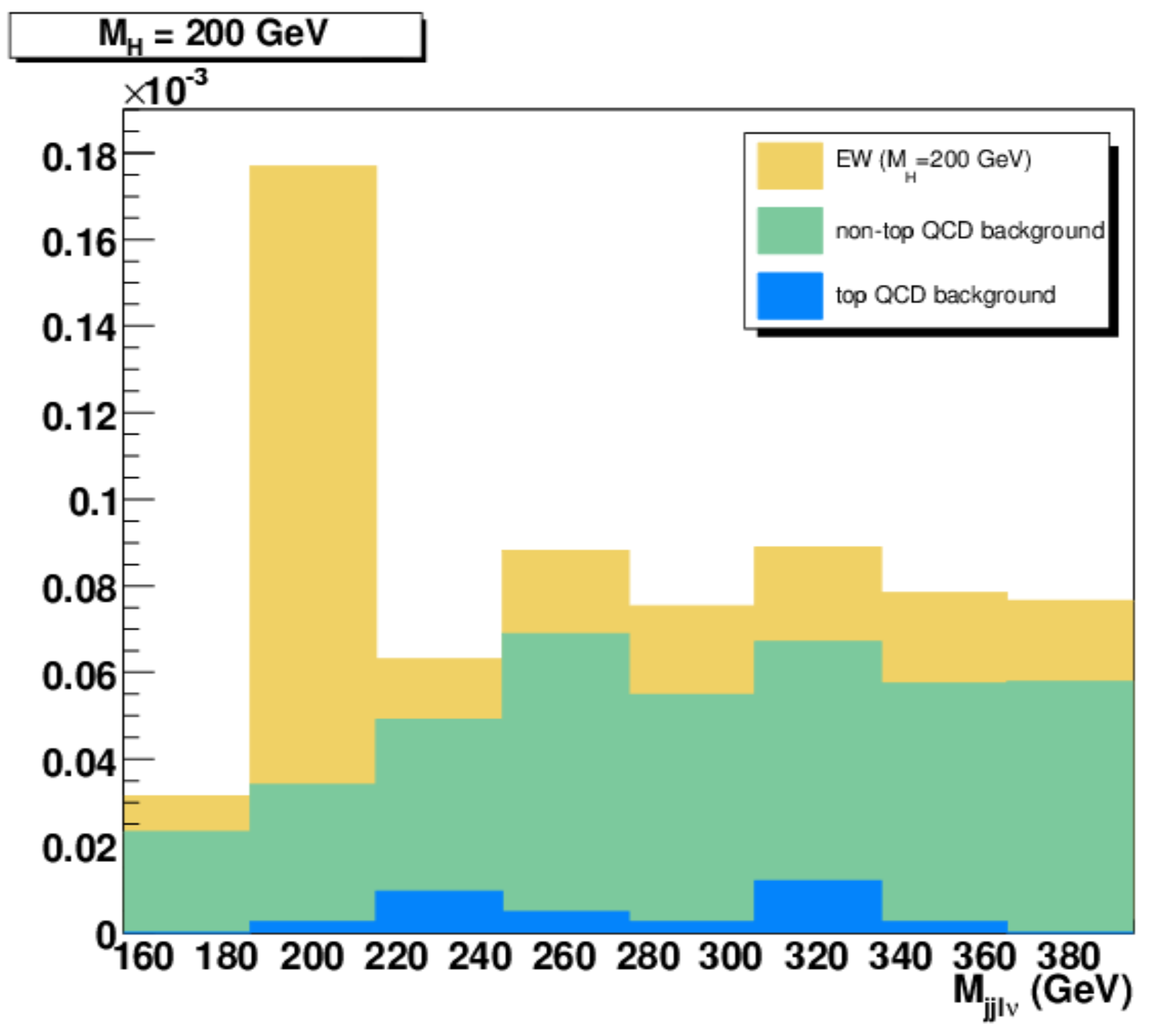}
\includegraphics*[width=0.49\textwidth,height=6.4cm]{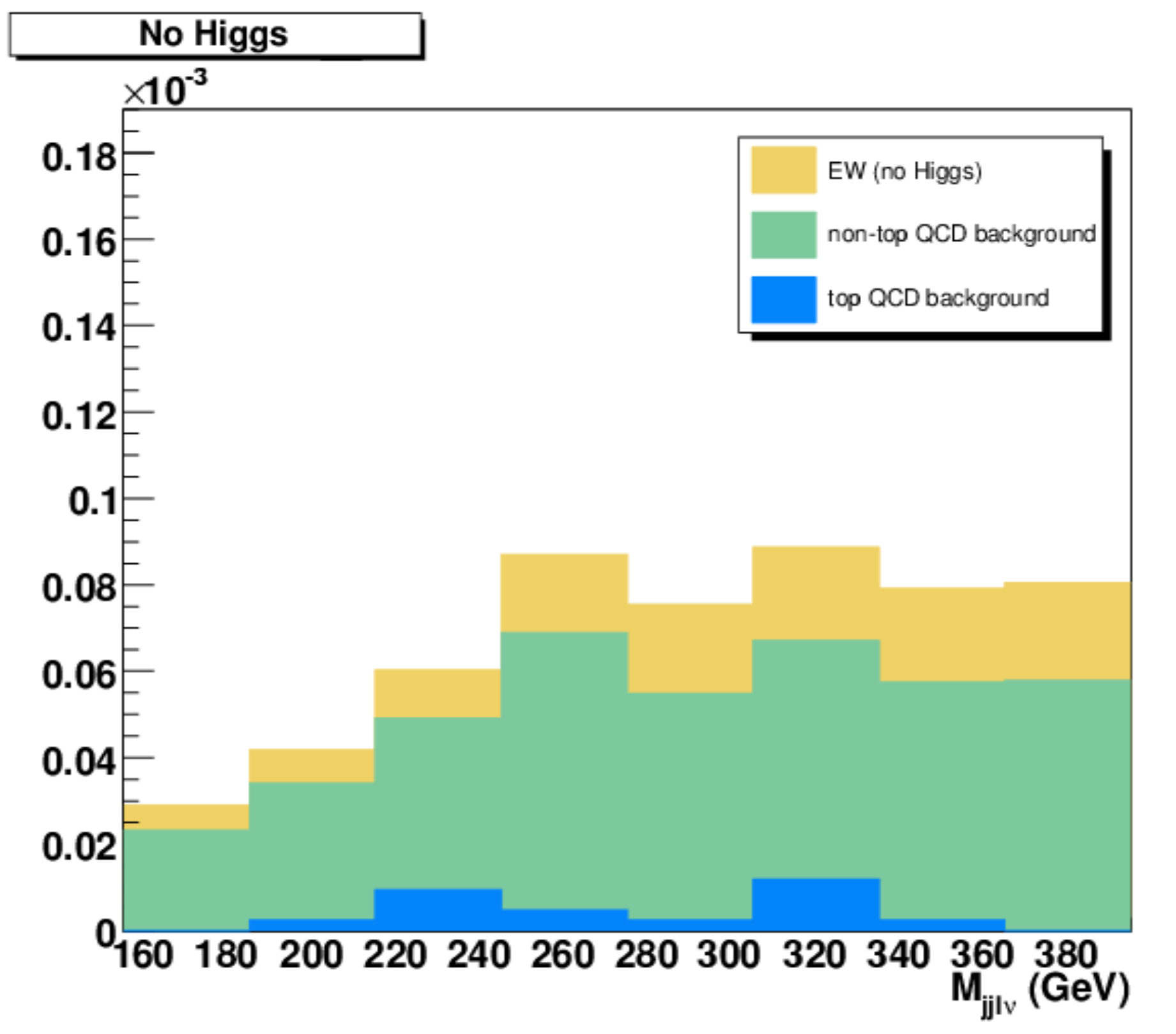}
\caption{ Invariant mass distribution of the two leptons and the two most 
central jets in the Standard Model with a light Higgs (on the left) and in the 
no-Higgs scenario (on the right). The cuts applied are listed in 
Tab.\ref{tab:cuts_munuqcd}.
\ordEW (EW) and \ordQCD (QCD) contributions to the differential cross 
section have been isolated and are shown separately. The QCD contributions are
further split into \textit{top background} (in blue) and \VVjj 
(in green).}
\label{fig:munuqcd_compare_signal_bckg}
\end{figure}

Fig.\ref{fig:munuqcd_compare_signal_bckg}
illustrates the final results of our analysis, showing that the top 
background is basically under control. \VVjj still provides a 
non-negligible contribution over the whole invariant mass spectrum, 
nevertheless differences between light-Higgs and no-Higgs can be appreciated.

In Tab.\ref{tab:xsec_EW},\ref{tab:xsec_EW+QCD} we show the integrated 
cross section at high energies as a function of the minimum invariant mass,
comparing results for the pure EW and the EW+QCD cases.
Despite reducing the ratio between no-Higgs and light-Higgs cross sections, the
inclusion of QCD background does not seem to compromise the possibility of
detecting signals of EWSB at the LHC from an excess of scattering events.
We find that about 500 events are expected above 800 GeV after one year of high
luminosity running ($\mathcal{L}=100\mbox{ fb}^{-1}$) in case of a Higgs boson 
with mass 200 GeV. The no-Higgs model predicts about 250 more events in 
accordance with the enhancement of the \VV differential cross section at high 
energies. These numbers refer to the muon channel only, and can obviously
be improved by summing up the muon and electron channels.
It should nevertheless be noticed that imposing a minimum $\Delta R$ separation
among colored particles could degrade these preliminary results and requires 
further investigations.

\section{Conclusions}
We have described in detail the features of the  Monte Carlo 
\Phantom which is the first dedicated event generator for six fermion physics which
 can be used at high energy  $pp$, $p\overline{p}$ and $e^+e^-$ colliders. 
It has already been used for
some phenomenological studies on boson boson scattering physics at the LHC and a short
example of its possibilities in this field has been presented. It will be used for
further detailed analyses at parton level, which should also contribute to assess the 
complementary role of hadronic and $e^+e^-$  colliders in this field.
It is however important to stress that after partonic studies, more  realistic 
ones with full event reconstruction and detector simulation are needed to fully
exploit the physics possibilities.  \Phantom has been developed also as a tool 
for these more complete analyses.

The program can be downloaded from http://www.ph.unito.it/{\small $\sim$}ballestr/phantom
where all new versions will be made available.

\section *{Acknowledgments}
A.B. wishes to thank the Department of Theoretical Physics of Torino University for support.

This work has been supported by MIUR under contract 2006020509\_004 and by the
European Community's Marie-Curie Research 
Training Network under contract MRTN-CT-2006-035505 `Tools and Precision
Calculations for Physics Discoveries at Colliders'

\newpage

\bc
{\bf \Large {Appendices}}
\ec
\appendix

\section{Routines}
\label{sec:routines}

\noindent Steering routines.\noindent The first one is the main program, 
the second drives the integration. The third, in generation mode, reads 
all grid files and steers the unweighted sampling.
\begin{verbatim}
   phantom.f   integ.f   oneshot.f
\end{verbatim}
Routine which parses the input file.
\begin{verbatim}
   readinput.f 
\end{verbatim}
As the name indicates {\tt coupling.f} computes the EW couplings in the
$G_\mu$-scheme, set masses and computes decay widths.
\begin{verbatim}
   coupling.f
\end{verbatim}
Modified \veg package.
\begin{verbatim}
   phavegas.f
\end{verbatim}
Integration function for {\t phavegas.f}
\begin{verbatim}
   fxn.f
\end{verbatim}
These routines select the set of channels to be used for each process and
initialize the corresponding parameters.
\begin{verbatim}
   proc.f   procini.f   procextraini.f
\end{verbatim}
Phase space routines. The {\tt XYZ\_jac.f} routines, not shown here, compute the
jacobian for the corresponding channel from the momenta.
\begin{verbatim}
   phsp1_1_31_multi_c.f     phsp1_1_4_multi5_c.f  
   phsp1_2_3_multi5_c.f     phsp1_5to1_4to31_multi_c.f 
   phsp2_4_multi5.f         phsp2_4to31_multi5.f 
   phsp3_3_multi5.f
\end{verbatim}
Beamstrahlung\cite{beam} and Initial State Radiation routines for  \epem colliders.  
\begin{verbatim}
   circe.f     isr.f 
\end{verbatim}
Implementing the cuts
\begin{verbatim}
   cuts.f
\end{verbatim}
These routines compute the complete amplitude for each process
using the master amplitudes  and the color evaluation routines mentioned  below.
\begin{verbatim}
   ampem.f   amp8fqcd.f   amp2g.f  
\end{verbatim}
Master amplitudes: 4W, 4Z, 2W2Z, 2g3Z, ggW2Z.
\begin{verbatim}
   fourw.f      fourz.f      twoztwow.f    
   fourwqcd.f   fourzqcd.f   twoztwowqcd.f
   gg3z.f       ggzww.f
\end{verbatim}
Color evaluation routines
\begin{verbatim}
   colevalew.f   coleval.f
\end{verbatim}
Storing the event information according to the latest Les Houches Accord
\begin{verbatim}
   LHAFileInit.f   storeLH.f
\end{verbatim}
Utility routines
\begin{verbatim}
   phread.f    bernoul.f   util.f     ccfcsym.f   
   extrema.f   isign.f     perm2g.f   perm.f
\end{verbatim}

\section{Sample input file}
\label{sec:r.in}

We report here a sample imput file to integrate the process $\bar b\, \bar b \rightarrow 
\bar b\, \bar b\, \mu^+ \mu^-\, \nu_\mu\, \bar\nu_\mu$. The lines starting with * and
the text following a ! are comments which can be left without problem in the input file.
One can use this same file to generate an unweighted sample with the same cuts and options,
just changing {\tt ionesh  0} in {\tt ionesh  1} and fixing at the end the exact number of 
files {\tt nfiles} and their names for the processes to be generated. The  line containing
the indication of the process ({\tt iproc}) will in this case be ignored.
It is evident from the comments that some parts of the input are ignored when {\tt ionesh=0}
and some others when {\tt ionesh=1}.
We also recall that our convention is that whenever there is a {\tt yes/no} option
1 corresponds to yes and 0 to no. The {\tt EVENTUAL MORE RESTRICTIVE CUTS}
are not all reported for brevity.

\vsk 
\begin {small}
\begin{verbatim}
iproc    -5 -5 -5 -5 13 -13 14 -14

idum  -123456789   !random number seed must be a large negative number

PDFname   /home/phantom/phantom/lhapdf-5.2.2/PDFsets/cteq5l.LHgrid

*  i_PDFscale selects the PDF scale:
       ! =1 for all processes, based on pT's of ALL OUTGOING PARTICLES
       ! =2 process by process, based on pT of the (RECONSTRUCTED) TOP
       !    if possible, otherwise as done in option 1
i_PDFscale  2

*   i_coll  determines the type of accelerator:
*    1 => p-p   2 => p-pbar   3 => e+e-
i_coll  1

i_isr   0         ! yes/no ISR for e+e- collider only

i_beamstrahlung   0     ! yes/no beamstrahlung for e+e- collider only

* perturbativeorder = 1 alpha_em^6 with dedicated amp
*    2 alpha_s^2alpha_em^4   3 alpha_em^6 + alpha_s^2alpha_em^4
perturbativeorder 3

ionesh  0   ! 0= one process run, 1= one shot generation
 
ecoll    14000.d0    ! collider energy

rmh      -500.d0   ! Higgs mass (GeV). If negative Higgs diagrams are not
                   ! computed: equivalent to rmh=infinity.

i_ccfam  1           ! yes/no family+CC conjugate

* CUTS

i_e_min_lep    1         ! yes/no lepton energy lower cuts (GeV)
e_min_lep      20.d0

i_pt_min_lep   1       ! yes/no lepton pt lower cuts (GeV)
pt_min_lep     10.d0

i_eta_max_onelep 0   ! yes/no at least one lepton in absval of 
                     ! eta_max_onelep. If no leptons are present 
                     ! in the final state cut is ignored 
eta_max_onelep    3.d0

i_eta_max_lep  1       ! yes/no lepton rapidity upper cuts
eta_max_lep    3.d0

i_ptmiss_min  0        ! yes/no missing pt lower cuts 
ptmiss_min    50.d0

i_e_min_j     1        ! yes/no jet energy lower cuts (GeV)
e_min_j       20.d0

i_pt_min_j    1       ! yes/no jet pt lower cuts (GeV) 
pt_min_j      10.d0

i_eta_max_j   1       ! yes/no jet rapidity upper cuts   
eta_max_j     6.5d0

i_eta_jf_jb_jc   0    ! yes/no the following 3 cuts
eta_def_jf_min 1.d0   ! min rapidity for a jet to be called forward
eta_def_jb_max -1.d0  ! max rapidity for a jet to be called backward
eta_def_jc_max 3.d0   ! max rapidity for a jet to be called central 
                      !       (absval)

i_pt_min_jcjc 0       ! pt lower cut on two centraljets (GeV)
pt_min_jcjc 50.d0

i_rm_min_jj 1      ! yes/no minimum invariant mass  between jets (GeV)  
rm_min_jj 20.d0

i_rm_min_ll 1     ! yes/no minimum invariant mass between charged lept.
rm_min_ll 20.d0

i_rm_min_jlep 0  ! yes/no minimum invariant mass between jets and lepton 
rm_min_jlep 30.d0

i_rm_min_jcjc 0     ! yes/no minimum invariant mass between central jets 
rm_min_jcjc 20.d0
 
i_rm_max_jcjc 0     ! yes/no maximum invariant mass between central jets 
rm_max_jcjc 14000.d0

i_rm_min_jfjb  0   ! yes/no minimum invariant mass between forward 
                   !and backward jets
rm_min_jfjb 100.d0

i_eta_min_jfjb 0  ! yes/no minimum rapidity difference between forward 
                  ! and backward jet 
eta_min_jfjb 3.d0

i_d_ar_jj 0  ! yes/no minimum delta_R separation  between jets
d_ar_jj 0.7d0

i_d_ar_jlep 0  !yes/no minimum delta_R separation between jets and lepton
d_ar_jlep 0.7d0

i_thetamin_jj 0  ! yes/no minimum angle separation between jets (degrees)
thetamin_jj    15.d0

i_thetamin_jlep 0  ! yes/no minimum angle separation between jets  
                   ! and lepton (degrees)
thetamin_jlep    15.d0

i_thetamin_leplep 0 ! yes/no minimum angle separation between charged 
                    ! leptons (cosine)
thetamin_leplep    15.d0

i_usercuts 0    ! yes/no (1/0) additional   user-defined cuts


****** IF (IONESH.EQ.0) THEN 

acc_therm     0.01d0      ! thermalization accuracy

ncall_therm   300000  3000000     ! thermalization calls per iteration
                    ! for the first 3 and for the remaining iterations
itmx_therm    5                   ! thermalization iterations

acc           0.005d0         ! integration accuracy

ncall         5000000           ! integration calls per iteration

itmx          5             ! integration iterations

iflat         1         ! yes/no flat event generation

****** ELSEIF (IONESH.EQ.1) THEN

scalemax  1.1d0             !scale factor for the maximum

nunwevts   20000     !  number of unweighted events to be produced

iwrite_event 1   ! yes/no momenta of flat events written in .dat files

iwrite_mothers 1   ! yes/no information about intermediate 
                   !  particles (mothers) in .dat files

i_exchincoming  1

*           EVENTUAL MORE RESTRICTIVE CUTS: 
*           SAME LIST AS BEFORE WITH "os" POSTPENDED

iextracuts  0  ! yes/no more restrictive cuts at generation level

i_e_min_lepos  0             ! lepton energy lower cuts (GeV)
e_min_lepos    30.d0
.
.
.
i_usercutsos  0

*  nfiles= number of files from which take the input for 
*          oneshot generation
nfiles   3    ! number of input files for oneshot generation
/home/phantom/phantom/process1/phavegas01.dat
/home/phantom/phantom/process1/phavegas02.dat
/home/phantom/phantom/process2/phavegas01.dat

\end{verbatim}
\end{small} 

\end{document}